\documentclass{aa}  
\usepackage{natbib}
\usepackage{graphicx}
\graphicspath{{Fig/}}
\usepackage[colorlinks=true, allcolors=blue]{hyperref}
\usepackage{txfonts}

\makeatletter
\renewcommand*\aa@pageof{, page \thepage{} of \pageref*{LastPage}}
\makeatother

\begin{document} 
\nolinenumbers
 \title{First detection of deuterated methylidyne (CD) in the interstellar medium}

   \author{Arshia M. Jacob$^{1,2}$, Karl M. Menten$^{2}$,  Friedrich Wyrowski$^{2}$ \and
   Olli Sipil\"{a}$^{3}$
          }

   \institute{William H. Miller III Department of Physics \& Astronomy, Johns Hopkins University, 3400 North Charles Street, Baltimore, MD 21218, USA 
   \and 
   Max-Planck-Institut f\"{u}r Radioastronomie, Auf dem H\"{u}gel 69, 53121 Bonn, Germany
   \and 
       Centre for Astrochemical Studies, Max-Planck-Institut f\"{u}r Extraterrestrische Physik, Gie{\ss}enbachstra{\ss}e 1, D-85748 Garching bei M\"{u}nchen, Germany
    }

   \date{Received October 13, 2022; accepted April 27, 2023}
  \titlerunning{First detection of CD in the interstellar medium}
   \authorrunning{Jacob et al. 2023}
 \abstract{While the abundance of elemental deuterium is relatively low (D/H\,$\sim\!$ a few $\times10^{-5}$), orders of magnitude higher D/H abundance ratios have been found for many interstellar molecules, enhanced by deuterium fractionation. In cold molecular clouds ($T< 20~$K), deuterium fractionation is driven by the H$_2$D$^+$ ion, whereas at higher temperatures ($T\geq20-$30~K) gas-phase deuteration is controlled by reactions with CH$_2$D$^+$ and C$_2$HD$^+$. While the role of H$_2$D$^+$ in driving cold interstellar deuterium chemistry is well understood, thanks to observational constraints from direct measurements of H$_2$D$^+$, deuteration stemming from CH$_2$D$^+$ is far less understood as a result of the absence of direct observational constraints of its key ions. Therefore, making use of chemical surrogates is imperative in order to explore deuterium chemistry at intermediate temperatures. 
 Formed at an early stage of ion--molecule chemistry directly from the dissociative recombination of CH$_3^+$ (CH$_2$D$^+$), CH (CD) is an ideal tracer for investigating deuterium substitution initiated by reactions with CH$_2$D$^+$. This paper reports the first detection of CD in the interstellar medium (ISM), carried out using the APEX 12~m telescope toward the widely studied low-mass protostellar system IRAS\,16293$-$2422. Observed in absorption towards the envelope of the central protostar, the D/H ratio derived from the column densities of CD and CH is found to be $0.016 \pm 0.003$. 
 This is an order of magnitude lower than the values found for 
 other small molecules like C$_2$H and H$_2$CO observed in emission but whose formation, which is similar to that of CH, is also initiated via pathways involving warm deuterium chemistry. 
 Gas-phase chemical models reproducing the CD/CH abundance ratio suggest that it reflects `warm deuterium chemistry' (which ensues in moderately warm conditions of the ISM) and illustrates the potential use of the CD/CH ratio in constraining the gas temperatures of the envelope gas clouds it probes.
 }


 \maketitle
\nolinenumbers
\section{Introduction} \label{sec:intro}

In recent years, the complexities of deuterium chemistry have been increasingly explored  \citep[see][for a review]{Ceccarelli2014}, motivated by the detection of a large number of deuterated molecules and molecular ions showing a range of deuterium substitutions from singly deuterated species such as HD \citep{Lacour2005} to triply deuterated species such as ND$_3$ \citep{Lis2002} and CD$_3$OH \citep{Parise2004}. Deuterium is a unique probe of primordial nucleosynthesis as it is formed in the very early Universe with an estimated abundance, D/H, of $\sim\!2.5 \times10^{-5}$ \citep{Cooke2018}. Being mainly destroyed in the interiors of stars via astration, the D/H ratio is expected to vary with location within the Galaxy. 
Remarkably, observations have revealed that the isotopologues of molecules in which one or more hydrogen atoms are substituted by deuterium atoms have relative abundances (with respect to their main counterparts) that are larger than the elemental D/H ratio, with values as high as 30\%--70\,\% \citep[as found, for example, for ND/NH toward IRAS\,16293$-$2422,][]{Bacmann2010}. Enhancements in the observed D/H ratio are not unexpected and result from isotopic fractionation, which favours the production of the deuterated isotope owing to small differences in the zero-point vibrational energies of hydrogen and deuterium. However, this 
alone cannot account for the large discrepancies between the observed and elemental D/H abundance ratio, and other chemical processes must be involved that depend on the environment in which these molecules exist. Therefore, the physical and chemical conditions that prevail in the regions where the molecules are formed also play an important role in enhancing the observed D/H ratio. This makes the degree of deuterium fractionation a strong indicator of both the chemistry and the physical conditions of the studied sources, and even their age \citep{Brunken2014}.

\begin{figure*}
    \sidecaption
    \includegraphics[width=12cm]{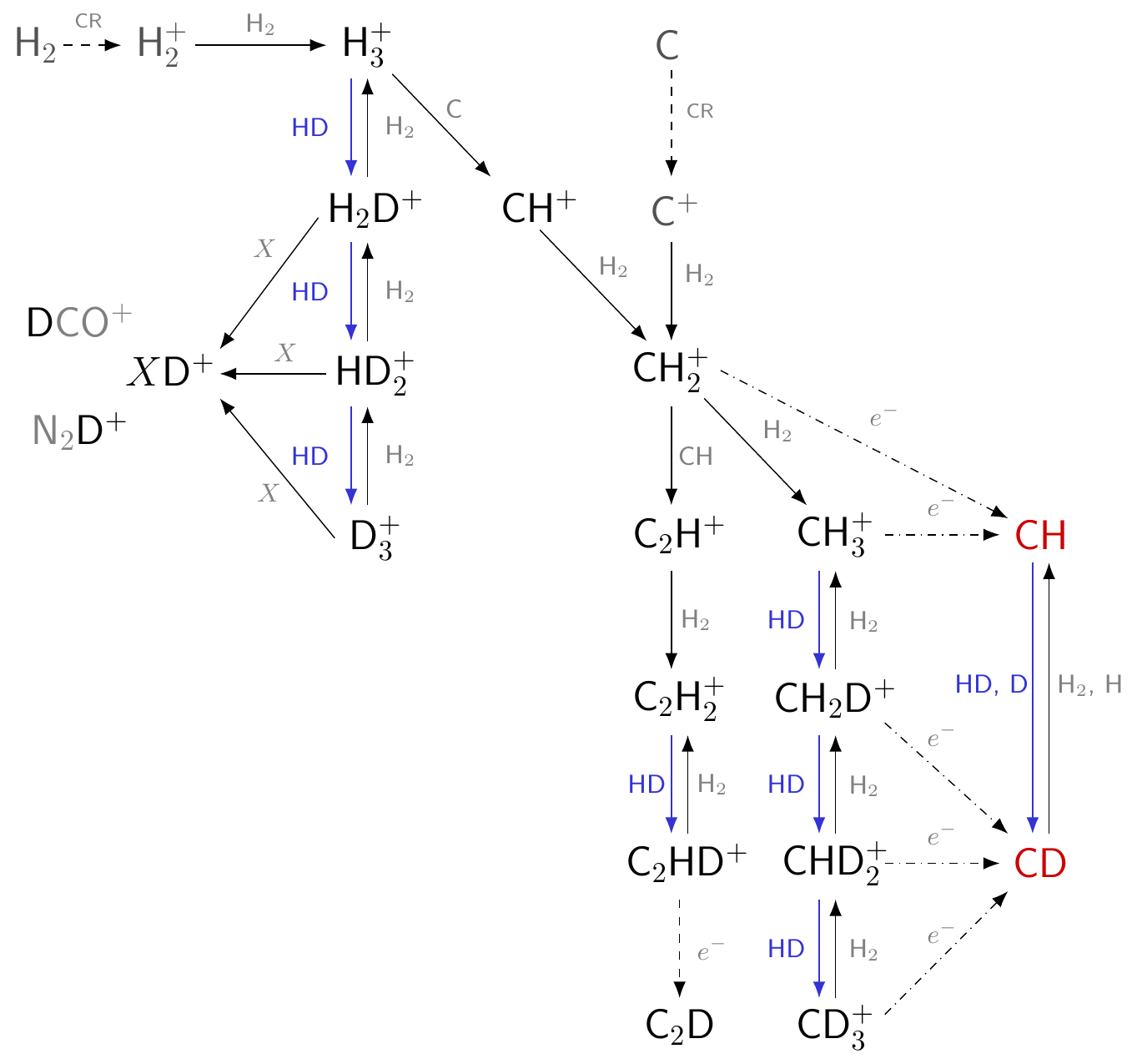}
    \caption{Network displaying the main reactions involved in deuterium fractionation, stemming from the H$_3^+$ cation. The relevant ion--neutral and neutral--neutral reactions alongside their corresponding reactants are marked using black arrows except for exothermic reactions involving HD or D, which are highlighted in blue. Ionisation by cosmic rays and electron recombination reactions are marked using dashed and dash-dotted black arrows, respectively.}
    \label{fig:network}
\end{figure*}

While D atoms are initially locked into HD, deuterium fractionation is initiated via fast ion--molecule reactions beginning with H$_{3}^+$  and forming H$_{2}$D$^+$, as shofwn in the left-hand side of Fig.~\ref{fig:network}. As the ensuing proton--deuteron exchange reaction is exoergic ($\Delta E= 232$~K), deuterium is enriched in the 
cold interstellar medium (ISM; $T\!\leq\!20$\,K), for example inside cold (${\approx\!10}$\,K) dark clouds, as the backward reaction (or hydrogenation) is inhibited. At higher gas temperatures (${T\!\sim20}$\,--\,80\,K), CH$_2$D$^+$ and C$_2$HD$^+$, formed via reactions of HD with CH$_{3}^+$ \citep{Asvany2004} and C$_2$H$_2^+$ \citep{Herbst1987}, are two other ions that effectively drive deuterium fractionation in the ISM. 
Furthermore, lying at the heart of carbon-chemistry, observations of CH$_3^+$ alongside CH$_2$D$^+$ and CHD$_2^+$, would present unique probes for studying deuterium chemistry in warmer environments of the ISM. For example, observed enhancements in the abundance of species like DCN in warm gas \citep{Leurini2006} have been interpreted to be a result of high CH$_2$D$^+$/CH$_3^+$ ratios \citep{Roueff2007}. However, lacking a permanent dipole moment owing to its symmetric and planar structure, CH$_3^+$ has no observable rotational transitions. Fortunately, this is not the case for its deuterated counterparts (CH$_2$D$^+$, CHD$_2^+$), which have rotational transitions observable at millimetre and radio wavelengths. However, to date, only tentative detections of two CH$_2$D$^+$ lines 
have been reported towards Orion IRc2 by \citet{Roueff2013}. Therefore, the successful detection of CD, which is formed from CH$_{2}$D$^+$, would be a strong indicator of both chemistry and the efficiency of secondary deuteration, particularly in the absence of secure detections of CH$_2$D$^+$.\\

One of the first molecules to be detected in the ISM, the methylidene radical, CH, first observed at visible wavelengths \citep{Dunham1937, Swings1937}, has been extensively studied owing to its importance in astrochemistry ---produced and destroyed at early stages in the sequence of ion--molecule reactions that govern interstellar chemistry--- and its use as a surrogate for H$_{2}$ in diffuse and translucent clouds \citep[e.g.][]{Federman1982, Sheffer2008, Weselak2010}. In contrast to CH, which has been observed in a wide range of astrophysical environments, across wavelengths ranging from the far-ultraviolet \citep[FUV;][]{Watson2001, Sheffer2007} to radio \citep[][]{Rydbeck1973}, very little is known about its rarer isotopologues, CD and $^{13}$CH. 
While considerable efforts have been made in the laboratory to measure the rotational spectra of these species \citep{Halfen2008, Zachwieja2012}, observationally, they have remained undetected in the ISM, until recently. \citet{Jacob2020} reported the first detection of $^{13}$CH towards four well-known star-forming regions in the Galaxy through observations of the hyperfine structure splitting components 
of the $N, J=2, 3/2 \rightarrow 1, 1/2$ transition of $^{13}$CH near ${\sim\!1997}~$GHz made possible  by the unique access of the GREAT\footnote{The German REceiver for Astronomy at Terahertz frequencies (GREAT) is a development by the MPI f\"{u}r Radioastronomie and the KOSMA/Universit\"{a}t zu K\"{o}ln, in cooperation with the DLR Institut f\"{u}r Optische Sensorsysteme.}\citep{Risacher2016} instrument on board the Stratospheric Observatory for Infrared Astronomy \citep[SOFIA;][]{Young2012} to supra-terahertz frequencies. In contrast, previous searches for the $N, J = 2, 3/2 \rightarrow 1, 3/2$ and $N, J = 2, 5/2 \rightarrow 1, 3/2$ transitions of CD near 885~GHz and 916~GHz made towards the low-mass protostar, IRAS 16293$-$2422\footnote{Well studied for its rich submillimetre spectrum and notably high degree of deuterium fractionation found for many molecules \citep{vanDishoeck1995}.}, by \citet{Bottinelli2014B} using the \textit{Herschel} space observatory were unsuccessful, leaving CD as yet undetected in the ISM.\\

This paper reports the first detection of CD in the ISM, towards the low-mass protostar IRAS\,16293$-$2422. Sections~\ref{sec:cd_spectroscopy} and \ref{sec:observations} briefly describe CD spectroscopy and the observations carried out, and our results are presented in Sect.~\ref{sec:results}. The chemical model used in the analysis and the subsequently derived constraints on deuterium chemistry are discussed in Sect.~\ref{sec:discussion}, with the main findings summarised in Sect.~\ref{sec:conclusions}. \\

\section{CD spectroscopy} \label{sec:cd_spectroscopy}
\begin{figure}[t]
    \centering
    \includegraphics[width=0.5\textwidth]{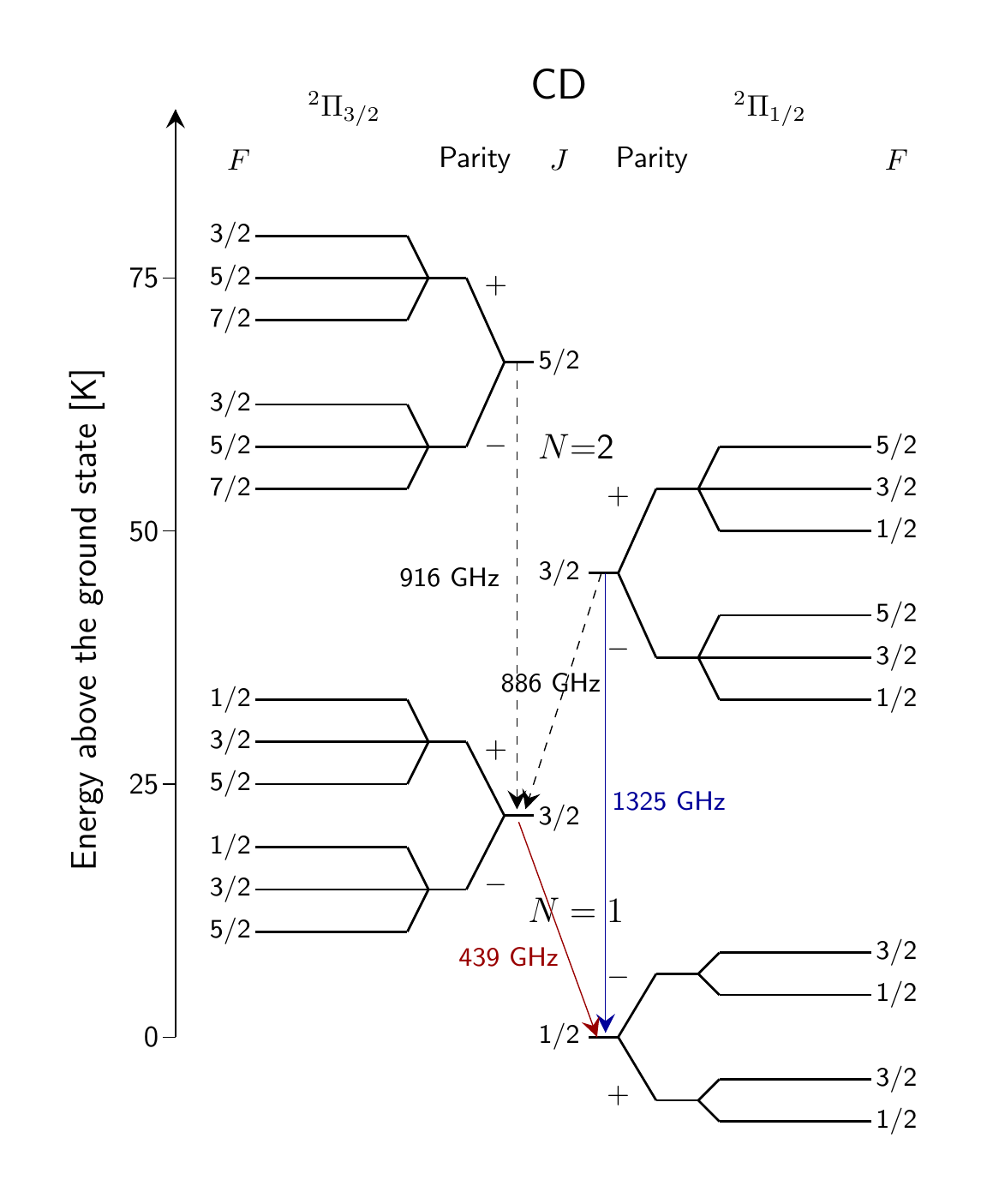}
    \caption{Energy-level diagram of the lowest rotational levels of CD, adapted from Figure 1 of \citet{Halfen2008}. The rotational transitions of CD discussed in the current study are highlighted using red and blue arrows. In addition, dashed black arrows indicate transitions of CD that were previously searched for, but remain undetected. We note that the level separations are not to scale.}
    \label{fig:eneregy_levels}
\end{figure}

The electronic, (ro-)vibrational, and purely rotational transitions in the $X^{2}\Pi_{r}$ ground electronic state of CD have been extensively studied using laboratory experiments. For example, the emission and absorption spectra of CD at visible and near-ultraviolet (NUV) wavelengths consisting of the $A^2\Delta$, $B^2\Sigma^{-},$ and $C^2\Sigma^{+}$ electronic systems were recorded first by \citet{Herzberg1969}, and more recently by \citet{Zachwieja2012}, \citet{Szajna2016}, and \citet{Para2018}, while the technique of high-resolution laser magnetic resonance (LMR) spectroscopy was used to investigate the rotational and vibrational transitions of this species at far-infrared (FIR) wavelengths by \citet{Brown1989}. CD, like CH and $^{13}$CH, has a $X^{2}\Pi_{r}$ ground state that conforms to a Hund's case (b) coupling, with each total angular momentum level $\boldsymbol{J}$ splitting into two parity states ($+$ and $-$) due to $\Lambda$-doubling. In addition, owing to the non-zero nuclear spin, $\boldsymbol{I}$, of deuterium, $I({\rm D}) = 1$, each rotational level further splits into hyperfine structure (HFS) levels, $\boldsymbol{F}$ ($=\boldsymbol{J+I}$). Using refined spectroscopic parameters (such as the rotational constant) derived by \citet{Wienkoop2003}, \citet{Halfen2008} were able to identify HFS splitting levels corresponding to the fundamental rotational transitions of CD in their laboratory spectra and in turn determine the rest frequencies of these transitions up to an accuracy of a few 100~kHz. The energy level diagram of the lowest rotational levels of CD, displaying the $\Lambda$-doublet transitions discussed in this work, is shown in Fig.~\ref{fig:eneregy_levels}, while the frequencies and spectroscopic parameters of the individual HFS transitions relevant for this work are summarised in Table~\ref{tab:spec_properties}. 


\begin{table}[t]
    \centering
    \caption{Spectroscopic parameters of the $N=1, J=3/2\rightarrow1/2$ transitions of CD.  }
    \begin{tabular}{lllcl}
    \hline \hline 
          \multicolumn{2}{c}{Transition}& \multicolumn{1}{c}{Frequency } & \multicolumn{1}{l}{$A_{\rm ul}\times10^{-3}$} \\
          Parity & $F^{\prime} - F^{\prime \prime}$ & \multicolumn{1}{c}{[MHz]}  &  \multicolumn{1}{c}{[s$^{-1}$]} \\
         \hline
            $+ \rightarrow -$ & $1/2 \rightarrow 3/2$ & ~439254.774(45) &   0.05 \\
                       & $3/2 \rightarrow 3/2$ & ~439255.608(30) &   0.22 \\
                       & $5/2 \rightarrow 3/2$\tablefootmark{*} &  ~439257.450(30) &  0.50 \\
                       & $1/2 \rightarrow 1/2$ & ~439271.905(30) &   0.45 \\
                       & $3/2 \rightarrow 1/2$ & ~439272.694(30) &  0.28 \\
     $- \rightarrow +$ & $5/2 \rightarrow 3/2$\tablefootmark{*} & ~439794.923(30) &  0.50 \\
                       & $3/2 \rightarrow 3/2$ & ~439800.005(30) &   0.22 \\
                       & $3/2 \rightarrow 1/2$ & ~439803.008(30) &  0.28 \\
                       & $1/2 \rightarrow 3/2$ & ~439803.124(45) &  0.05 \\
                       & $1/2 \rightarrow 1/2$ & ~439806.093(30) &  0.44 \\
                       \hline
  
\hline
    \end{tabular}
    \tablefoot{{The frequencies are taken from laboratory measurements by \citet{Halfen2008}, while the other spectroscopic parameters presented are taken from the Jet Propulsion Laboratory \citep[JPL;][]{Pickett1998} database. Values in parentheses represent uncertainties in the rest frequencies of the HFS lines, in units of the last significant digits.} Asterisks indicate the HFS splitting transition, which was used to set the velocity scale in the analysis.}
    \label{tab:spec_properties}
\end{table}

\section{Observations} \label{sec:observations}
We carried out a search for the $^{2}\Pi_{3/2}, N=1, J=3/2 \rightarrow ^{2}\Pi_{1/2}, N=1, J=1/2$ HFS splitting transitions of CD near 440~GHz using the high-frequency module of the new version of the First Light APEX Submillimetre Heterodyne Receiver, nFLASH460, a recently commissioned sideband separating (2SB) receiver on the APEX 12~m submillimetre telescope\footnote{APEX is a collaboration between the Max-Planck-Institut fur Radioastronomie, the European Southern Observatory, and the Onsala Space Observatory.} \citep{Gusten2006}. 

\begin{figure}
    \centering
\includegraphics[width=0.43\textwidth]{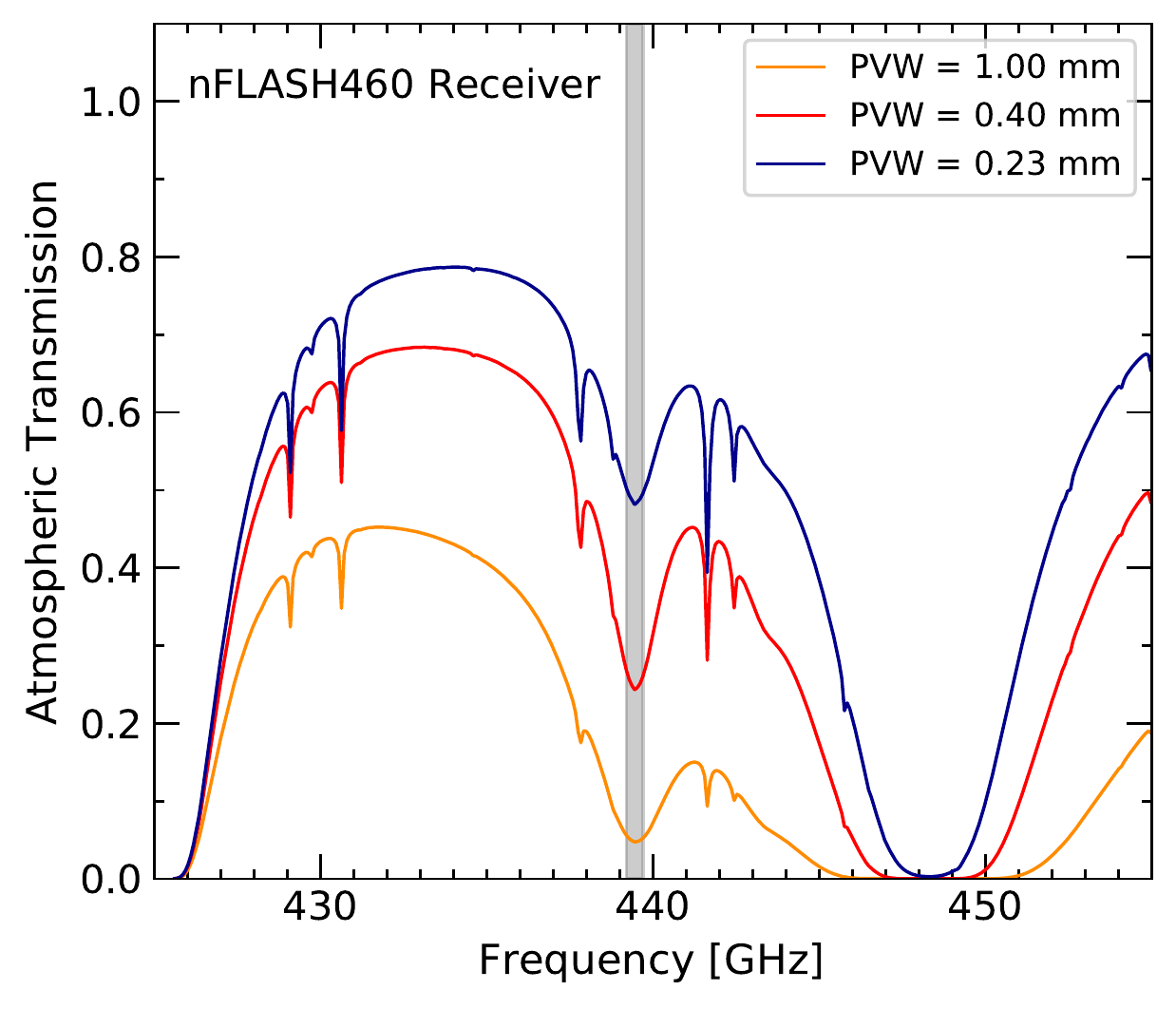}
    \caption{Calculated atmospheric zenith transmission of the 440~GHz ($\sim 680~\mu$m) submillimetre window for the site of the APEX telescope for three values of the precipitable water vapour column: 0.23~mm (dark blue), 0.40~mm (red), and 1.00~mm (orange). The grey shaded region marks the range of frequencies covered by all the HFS splitting transitions between $^{2}\Pi_{3/2}, N=1, J=3/2 \rightarrow ^{2}\Pi_{1/2}, N=1, J=1/2$ of CD.}
    \label{fig:atm_transmission}
\end{figure}

Observations of all of the HFS splitting components of the CD line are challenging due to the proximity of their frequencies to the $4_{23}-3_{30}$ line of ortho-H$_2$O at 448.0010775~GHz, and other H$_2$O lines even nearer in frequency. The line strength of the 448~GHz line causes significant atmospheric absorption with extended line wings \citep[see Fig. 1 of][]{Menten2008}. Referring to this figure, we note that our observations were made under outstanding weather conditions, when the precipitable water vapour column was between 0.23 and 0.45~mm (see Fig.~\ref{fig:atm_transmission}, where the atmospheric transmission curves are computed based on the \textit{am} transmission model\footnote{See \url{https://www.cfa.harvard.edu/~spaine/am/} for more information on the $am$ atmospheric model.} developed by the Smithsonian Receiver Lab
at the Smithsonian Astrophysical Observatory). 
The observations were carried out across four observing sessions in 2021 June and seven in 2022 April and June, respectively, under project id: M9530C\_107. The FFTS4G backend, which covers bandwidths of 4~GHz in each sideband, provided a spectral resolution of 61~kHz corresponding to a native velocity resolution of 0.041~km~s$^{-1}$. The half-power beam width (HPBW) is 14\rlap{.}$^{\prime\prime}$3 at 440~GHz, and a forward efficiency of 0.95 and a main beam efficiency of ${\sim\!0.53}$ were assumed. The data were subsequently reduced and processed using the GILDAS/CLASS software\footnote{Software package developed by IRAM, see \url{https://www.iram.fr/IRAMFR/GILDAS/} for more information regarding the GILDAS package.}\citep{Pety2005}. Polynomial baselines up to second order were removed and the resultant spectra were box-smoothed to channel widths
of 0.082~km~s$^{-1}$. The total integration time for each pointing is given in Table~\ref{tab:source_prop}. The observations were carried out in wobbler-switching mode using a throw of the wobbling secondary of 120$^{\prime\prime}$ in azimuth at a rate of 1.5~Hz, which is fast enough to reliably recover the continuum emission of the background sources. It should be noted that the online calibration takes into account variations in transmission close to the water lines with a channel-by-channel atmospheric calibration.
 \\

In this pilot study, we observed the low-mass protostar, IRAS\,16293$-$2422, and three well-known Galactic massive star-forming regions: Sgr\,B2\,(M), G34.26+0.15, and W49\,N.\\

Located within the Ophiuchus star-forming complex, IRAS\,16293$-$2422 has been the target of numerous astronomical investigations \citep[e.g.][]{Ceccarelli1998, Ceccarelli2001, Parise2002, Parise2004, Calcutt2018, Melosso2020}, in particular thanks to the high degrees of deuterium fractionation found for many molecules. Towards this source, the deuterated counterparts of several key species in astrochemistry were first detected, for example ND \citep{Bacmann2010}, OD \citep{Parise2012}, NHD, ND$_2$ \citep{Melosso2020}, and CH$_{3}$OCHD$_{2}$ \citep{Richard2021}. The protostellar system consists of two hot corinos, A and B, separated by a distance of 5$^{\prime\prime}$ in the plane of the sky \citep{Wootten1989, Mundy1992, Loinard2002} where source A further consists of subcondensations, A1 and A2 \citep{Maureira2022}. While the pointed observations are carried out towards a position closer to subcondensation A2 in core A, the different components are not resolved in the $\sim\!$14$^{\prime\prime}$ HPBW of the nFLASH460 receiver, which in fact also covers contributions from core B, as well. The massive star-forming regions selected for this study, Sgr\,B2\,(M), G34.26+015, and W49\,N, are all characterised by strong (sub)millimetre continua, towards which several deuterated species, both simple and complex, have been detected \citep{Comito2010, Fontani2011, Liu2013, Belloche2016, Zahorecz2021} with relatively high degrees of deuteration (e.g. \citet{Wienen2021} find deuterium enhancements for NH$_2$D ${\sim\!3}$\% even towards the hot core of G34.26+0.15). In addition, $^{13}$CH has been detected towards these star-forming regions \citep{Jacob2020}. Properties of the different sources as well as observational parameters, including the pointing positions used, are provided in Table~\ref{tab:source_prop}. \\

When a line is observed in absorption, it is essential to account for calibration uncertainties in the absolute continuum levels as the line-to-continuum ratio forms the crux of the analysis that follows. For our data, on average the continuum levels estimated from the line-free region of spectra vary by $<\!14\%$ across scans. Furthermore, for the massive star-forming regions, we find the 440~GHz continuum fluxes to be well correlated with the 870~$\mu$m continuum emission obtained in the course of the APEX Large survey of the Galaxy \citep[ATLASGAL;][]{Schuller2009}. The continuum flux at 440~GHz is also found to be consistent with that obtained from previous observations of these regions carried out using the Spectral and Photometric Imaging Receiver \citep[SPIRE;][]{Griffin2010} on the \textit{Herschel} Space Observatory (HSO).

In addition to the APEX observations discussed here, observations of the $N=2 \rightarrow 1, J=3/2 \rightarrow 1/2$ HFS splitting transitions of CD near 1325~GHz were carried out using the 4GREAT receiver \citep{Duran2021} on board SOFIA (project id: 83\_0701). The frequencies alongside other relevant spectroscopic parameters of these transitions are given in Table~\ref{tab:sofia_freq}. Due to their exploratory nature, and the limited time available on SOFIA, these observations were only carried out towards W49\,N. While the 4G3 module of GREAT was tuned to cover the frequencies of the CD transitions, the 4G1, 4G2, and 4G4 receivers were tuned to observe the $N_{K_{\rm a}K_{\rm c}} = 1_{1,0}-0_{0,0}$ and $N_{K_{\rm a}K_{\rm c}} = 1_{1,0}-0_{0,0}$ lines of HDO at 509.292~GHz and 893.638~GHz \citep{Johns1985} and the $N=3-2, J=5/2\rightarrow 3/2$ HFS splitting transitions of $^{18}$OH near 2494.690~GHz \citep{Beaudet1978}, respectively. Not pertaining to the results of this work, the spectra obtained from the other 4GREAT channels are not discussed any further here.

\begin{table*}
 \caption{Continuum source parameters.}
    \centering
    \begin{tabular}{lrrccccc}
    \hline \hline
    Source & \multicolumn{1}{c}{$\alpha$(J2000)} & \multicolumn{1}{c}{$\delta$(J2000)} & $\upsilon_{\rm LSR}$ & $T_{\rm c}$ & $T_{\rm rms}$ & $t_{\rm obs}$\\
    Designation & \multicolumn{1}{c}{[$^{\rm h}: ^{\rm m}: ^{\rm s}$]} & \multicolumn{1}{c}{[$^{\circ}:^{\prime}:^{\prime\prime}$]} & [km~s$^{-1}$]& [K] & [mK]& [hours]\\
    \hline
        IRAS\,16293$-$2422 & 16:32:22.78 & $-$24:28:38.70 & ~3.9 & 1.18 &  16 & 10.5\\
        Sgr\,B2\,(M) & 17:47:20.49 & $-$28:23:06.00 & 62.0 & 6.70 & 21 & ~1.9\\
        G34.26+0.15 & 18:53:18.60 &  01:14:58.00 &  57.0 & 3.05 & 12 & ~3.3\\
        W49\,N & 19:10:13.20 & 09:06:11.80 & 11.8 & 2.36 & 33 & ~2.6 \\
         \hline 
    \end{tabular}
   \tablefoot{$T_{\rm c}$ represents the continuum level at 440~GHz and $T_{\rm rms}$ the rms noise level on the $T_{\rm MB}$ scale quoted for a spectral resolution of 0.08~km~s$^{-1}$. The observing leg time ($t_{\rm obs}$) refers to the total duration of time for which each source was observed. }
    \label{tab:source_prop}
\end{table*}

\section{Results and analysis} \label{sec:results}
\begin{figure*}
    \centering
    \includegraphics[width=0.45\textwidth]{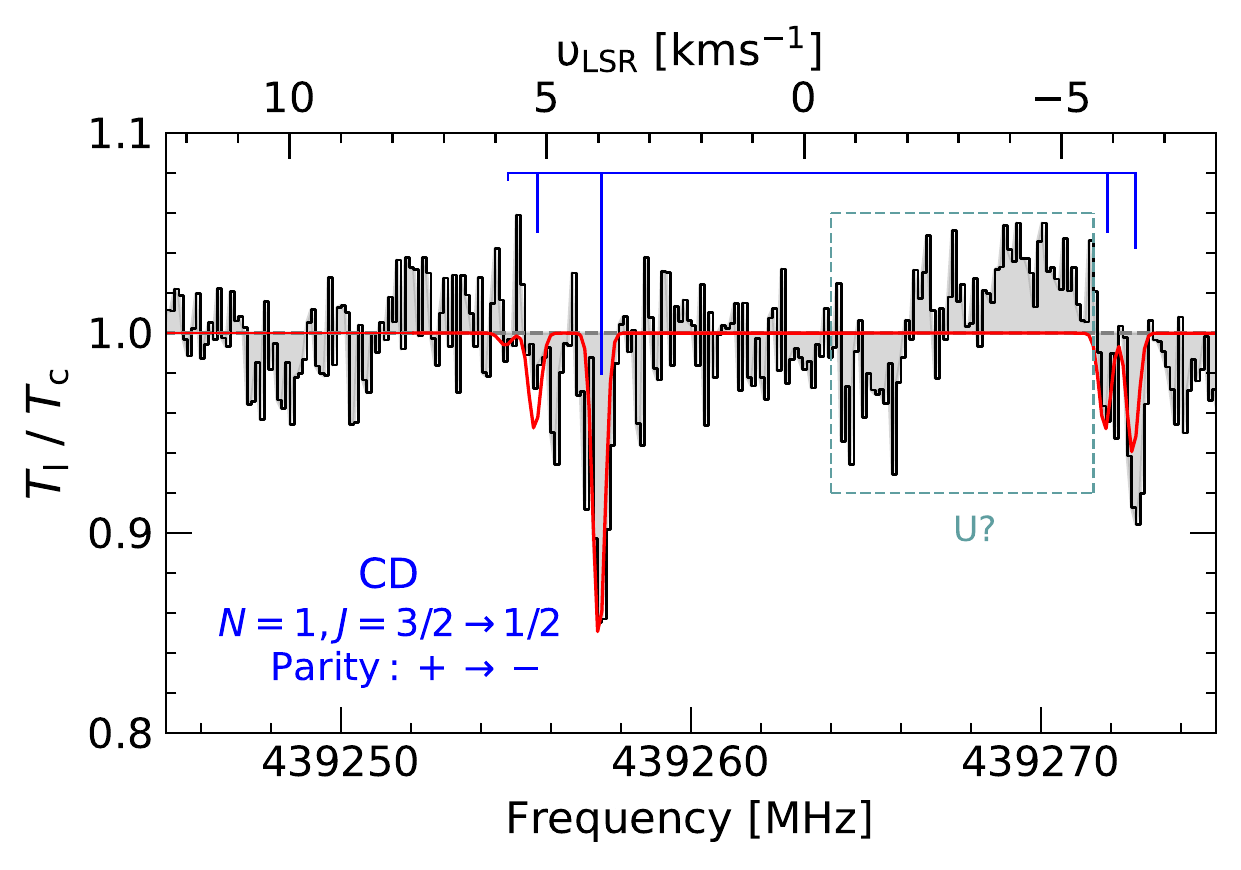}\quad 
    \includegraphics[width=0.464\textwidth]{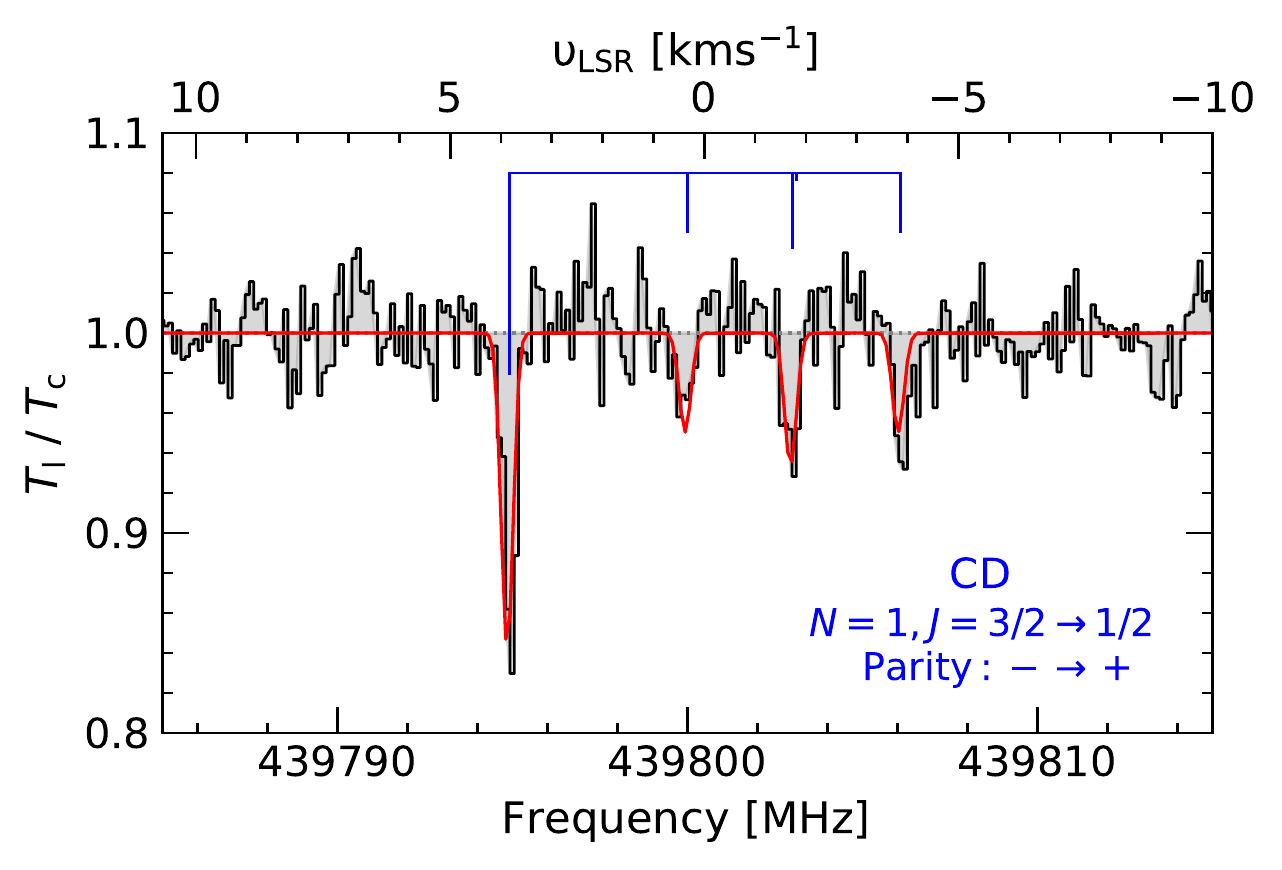}\\
    \includegraphics[width=0.45\textwidth]{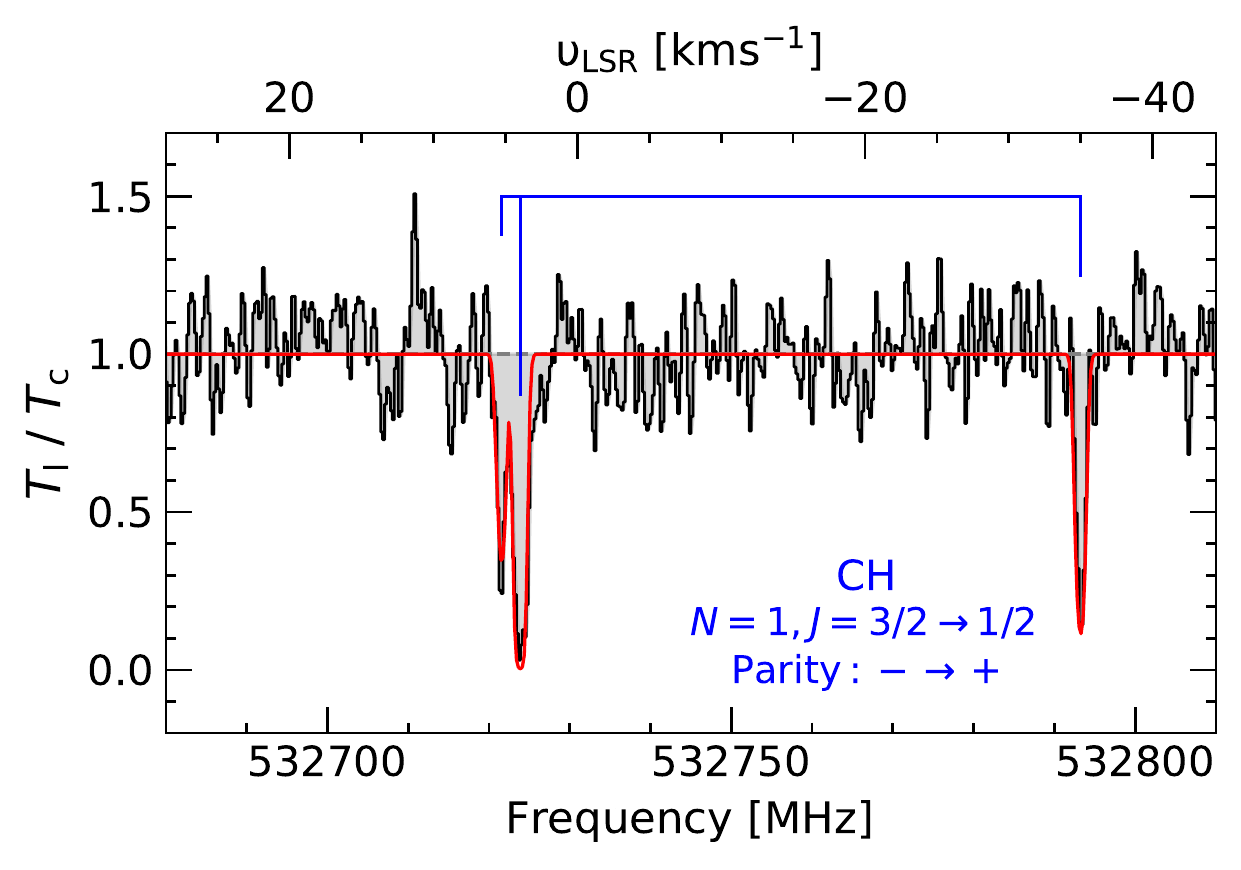}\quad
    \includegraphics[width=0.46\textwidth]{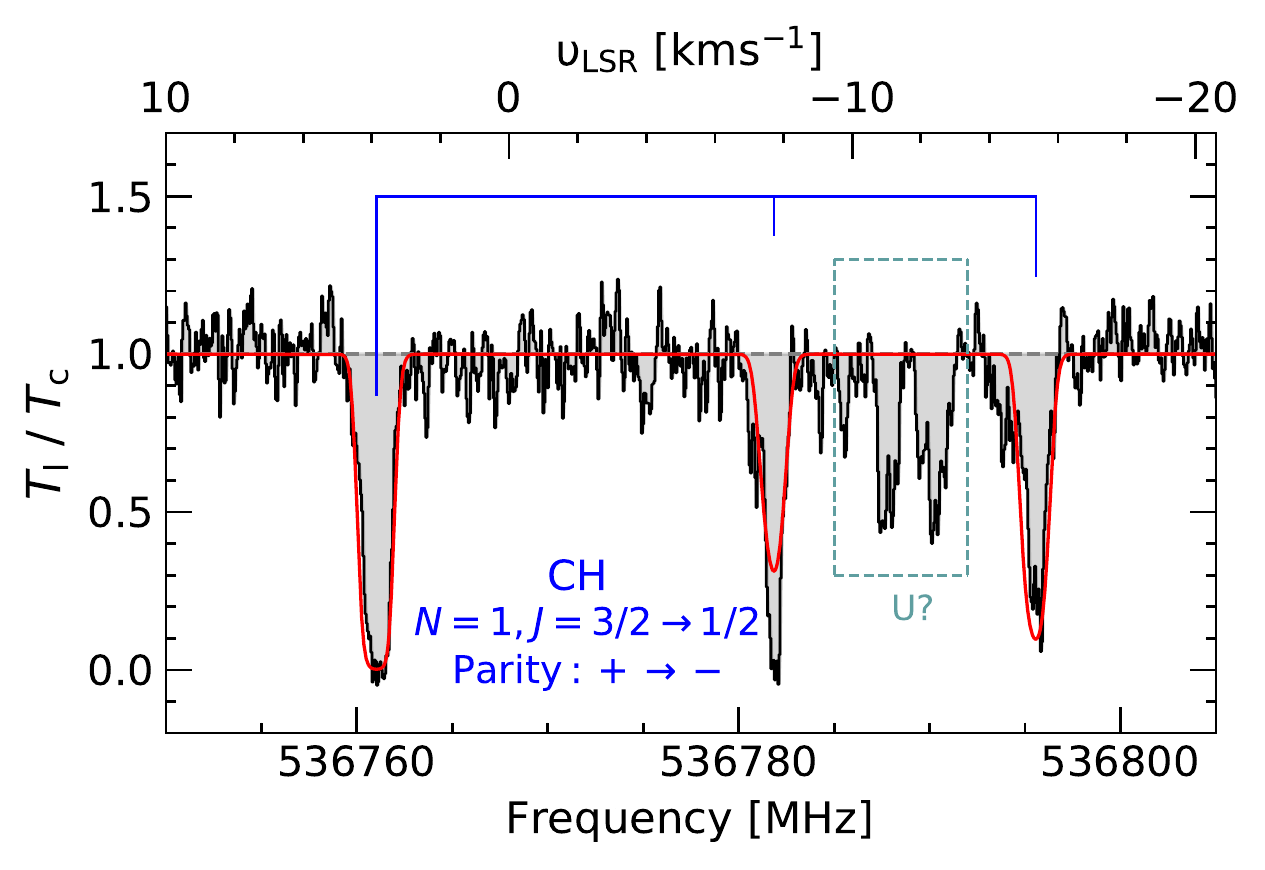}
    \caption{Spectra of the $N=1, J = 3/2\rightarrow 1/2$ transitions of CD and CH. Top panel: Filled grey histograms represent the normalised spectra of CD at 439260~MHz (left) and 439800~MHz (right) observed towards IRAS\,16293$-$2422 alongside the GILDAS-Weeds model fit overlaid in red. The relative intensities of the HFS splitting components are marked in blue. Unidentified, contaminating spectroscopic and/or baseline features are labelled- U? in light blue. Bottom panel: Same as the top panel but displaying the corresponding CH spectra at 532720~MHz (left) and 53680~MHz (right), respectively.}
    \label{fig:CD_spec}
\end{figure*}

Figure~\ref{fig:CD_spec} displays the detected HFS splitting lines of both $\Lambda$-doublets of the ${N=1, J=3/2 \rightarrow 1/2}$ transitions of CD towards IRAS\,16293$-$2422. The observed HFS lines of CD are seen in absorption with a centroid velocity of $3.9$~km~s$^{-1}$ \citep[consistent with the velocities at which other envelope gas tracers like o-ND$_2$ absorb,][]{Melosso2020}, narrow line widths of 0.25~km~s$^{-1}$ , and relative intensities close to what is expected under conditions of local thermodynamic equilibrium (LTE). This is not surprising as the observed line-to-continuum ratios for CD translate to very low optical depths with values between 0.04 and 0.15. The $F=1/2\rightarrow1/2$ and $F=3/2\rightarrow1/2$ HFS components of CD near 439272~MHz lie close to a spurious feature whose frequency does not correspond to any known transition recorded in molecular line databases. This, in addition to its comparatively broader line width, suggests that this feature is likely a spectral artefact and has no astronomical origin. 

For comparison, Fig.~\ref{fig:CD_spec} also displays the corresponding HFS resolved $\Lambda$-doublet spectra of CH near 532~GHz and 536~GHz (frequencies summarised in Table~\ref{tab:ch_freq}), respectively, taken from the \textit{Herschel} Science archive\footnote{\href{http://archives.esac.esa.int/hsa/whsa/}{http://archives.esac.esa.int/hsa/whsa/}} and discussed in \citet{Bottinelli2014B}. Dedicated pointing observations were carried out towards IRAS\,16293$-$2422 using the high-resolution spectrometer (HRS) on \textit{Herschel}/HIFI providing a spectral resolution of ${\sim\!0.07}$~km~s$^{-1}$ at 536~GHz (observation id: 1342214339) and $0.27$~km~s$^{-1}$ at 532~GHz (observation id: 1342227403, 1342227404). In addition, the frequencies of the CH lines were also covered as a part of the HIFI guaranteed time key program Chemical HErschel Surveys of Star-forming regions \citep[CHESS;][]{Ceccarelli2010} in double beam switch mode (observation id: 1342191499) with a spectral resolution of $\sim\!$0.6~km~s$^{-1}$ at 532/536~GHz. While the HRS observations provide a better spectral line resolution, the spectra are contaminated by absorption line features arising from the image sideband. 
Therefore, both sets of observations are essential to accurately identify contaminating absorption line features \citep[as detailed in Appendix A of][]{Bottinelli2014B} and to subsequently reveal the true depth of the CH absorption line profiles. We note that the position towards which the \textit{Herschel}/HIFI pointed observations were carried out is offset (in declination by 5$^{\prime\prime}$) from that of the APEX observations. However, the large HPBW of $\sim$\,38$^{\prime\prime}$ of HIFI band 1a at 530~GHz also ensures that both facilities sample the same gas.

The spectral lines are modelled using an extension of the GILDAS-CLASS software, Weeds \citep{Maret2011}, which is used to reduce and analyse spectral line observations under conditions of LTE, assuming that the excitation is dominated by the cosmic microwave background radiation, $T_{\rm CMB}=2.73~$K, which we assume to be the rotation temperature. This is a valid assumption given that the critical density of the two-level CH system studied, $n_{\rm crit} \geq 10^{6}\,$cm$^{-3}$ , is orders of magnitude higher than the ambient gas density in the envelope of IRAS\,16293$-$2422, in which the absorption arises in gas layers with $n = 10^{3}$--$10^{4}\,$cm$^{-3}$ \citep{Brunken2014}. As collisional rate coefficients are not presently available for CD, we assume the critical density of the CD transitions studied to be identical to that of the corresponding CH transitions. Furthermore, this rotation temperature allows a satisfactory fit to both the CD and CH absorption spectra, while at higher rotational temperatures the modelled CH absorption features begin to saturate. 
Following \citet{Bottinelli2014B}, we fitted the CH spectra using two components: the first corresponds to the envelope gas at ${\upsilon_{\rm LSR}\!\sim\!3.9}$~km~s$^{-1}$ with a line width of 0.57~km~s$^{-1}$ and the second to a foreground cloud at 4.2~km~s$^{-1}$ with a line width of 0.46~km~s$^{-1}$, a component that has been observed in the spectra of several other species \citep[e.g. HDO;][]{Coutens2012} towards this source. Given that the CD absorption is generally weaker than that of CH, we do not detect significant CD absorption from the foreground cloud at 4.2~km~s$^{-1}$. Adapting the formalism presented, for example, in Sect. 4 of \citet{Jacob2020} with the Weeds model as input, we calculate CD and CH column densities from the line-to-continuum ratios determined for the envelope of IRAS\,16293$-$2422. We derive a CD column density of (1.25$\pm$0.31)$\times10^{12}$\,cm$^{-2}$ and a CH column density of (7.71$\pm$0.25)$\times10^{13}$\,cm~$^{-2}$ \citep[similar to that derived by][]{Bottinelli2014B}. 
Assuming that the CD and CH distributions are of equal extent ---one that matches the continuum emission of the source--- results in a D/H ratio of 0.016$\pm$0.003. 
\begin{table*}[]
\caption{Synopsis of the derived CH and CD column densities and D/H ratios.}
    \centering
    \begin{tabular}{ll rl lrlll}
    \hline \hline 
         Source & $\upsilon_{\rm  min}$--$\upsilon_{\rm  max}$ & \multicolumn{1}{c}{$N$(CH)} & \multicolumn{1}{c}{$N$(CD)} & \multicolumn{5}{c}{D/H ratio}  \\
         &  \multicolumn{1}{c}{[km~s$^{-1}$]}
         &  \multicolumn{1}{c}{[10$^{13}$ cm$^{-2}$]} & \multicolumn{1}{c}{[10$^{12}$ cm$^{-2}$]} & \multicolumn{1}{c}{CH} &  \multicolumn{1}{c}{HCO$^+$} &\multicolumn{1}{c}{C$_2$H} &\multicolumn{1}{c}{H$_2$CO} &\multicolumn{1}{c}{HCN}\\
         \hline 
         IRAS~16293$-$2422 & \multicolumn{1}{c}{3.2 -- 4.7}& $7.71\pm0.25$ & $1.25\pm0.31$ & $0.016\pm0.003$ & $0.0086$\tablefootmark{a} & 0.18\tablefootmark{a} & 0.14\tablefootmark{a} & 0.013\tablefootmark{a}\\
         W49~N & $-5.0$ -- 25.0 & $22.87\pm3.60$& \multicolumn{1}{c}{$<6.85$}& \multicolumn{1}{c}{$\leq0.03$} & $\leq0.0004$\tablefootmark{b} & \multicolumn{1}{c}{--} & \multicolumn{1}{c}{--} & 0.001\tablefootmark{b}\\
         \hline
    \end{tabular}
    \tablefoot{\tablefoottext{a}{Values taken from \citet{vanDishoeck1995}.} \tablefoottext{b}{Values taken from \citet{Roberts2011}.}}
    \label{tab:coldens_comparison}
\end{table*}
Fortuitously, this value is comparable to that predicted by early chemical models of deuterated chemistry described by \citet{Millar1989} towards dense clouds. 


Towards the high-mass star-forming regions studied in this work, none of the observed transitions of CD are detected above noise levels between 12\,mK and 33\,mK at a velocity resolution of ${\sim\!0.1}$\,km~s$^{-1}$ and continuum levels of 6.7\,K, 3.05\,K, and 2.36\,K towards Sgr\,B2\,(M), G34.26+0.15, and W49\,N, respectively. Moreover, for the observed  hot cores, the 440~GHz spectral line window covers a forest of emission lines, which further complicates the identification of weak CD absorption features. In addition, estimating reasonable limits on the $N$(CD)/$N$(CH) abundance ratios is made difficult by the complex nature of the observed spectral line profiles of CH at 532/536~GHz, which show both emission and absorption features at the velocities of the background sources. 
The 1325~GHz lines of CD observed using SOFIA/4GREAT also remain undetected towards W49\,N, down to a noise level of 0.55\,K (at a velocity channel width of ${\sim\!0.1}$\,km~s$^{-1}$); however, unlike the CH lines at 532/536~GHz, the spectrum of the corresponding CH lines near 2007~GHz is seen in deep absorption unaffected by contaminating emission features \citep{Wiesemeyer2018}, thereby yielding reliable column density measurements. Combined with 3$\sigma$ level upper limits on the CD column density computed over the known velocity dispersion of the source in absorption, we report a $N$(CD)/$N$(CH) abundance ratio of $\leq\!0.03$. This is roughly a factor of two greater than the value derived towards IRAS\,16293$-$2422. The column densities of CH and CD determined towards IRAS~16293$-$2422 and W49~N are presented in
Table~\ref{tab:coldens_comparison} along with a comparison of the D/H ratio derived from CH with that of other chemical species.\\

\section{Discussion} \label{sec:discussion}

The degree of deuteration derived from the CD and CH column densities towards IRAS\,16293$-$2422 is comparable to that of HCO$^+$ and HCN but is an order of magnitude lower than the values derived from other small molecules such as C$_2$H and H$_2$CO \citep{vanDishoeck1995, Caux2011} ---molecules that are synthesised in the gas phase via reaction channels involving warm deuterium chemistry, towards the same source. The lower levels of deuteration derived for CH when compared to species like C$_2$H and H$_2$CO can be attributed to a combination of chemical effects, which may either enhance the production of CH or moderate the production of CD. Firstly, both CH and CD, observed in absorption, likely trace gas layers that may not be warm enough for deuterium fractionation reactions involving CH$_2$D$^+$ to be the sole reaction channel driving deuteration, unlike in the case of C$_2$H and H$_2$CO and their deuterated counterparts discussed above, which are observed in emission. In addition, CD is not the primary product of the dissociative recombination (DR) of CH$_2$D$^+$ with electrons. Analogous to CH$_3^+$, CH$_2$D$^+$ ions also combine with electrons to form different channels of ion--neutral products, including CHD, CH$_2$, CD, CH, and C.

While laboratory measurements have revealed the branching ratio of the DR of CH$_3^+$ to be ${35\!:\!30\!:\!35}$ between CH$_2$:CH:C \citep{Vejby1997, Thomas2012}, we lack similar measurements for the branching ratios of the CH$_2$D$^+$ by-products. Nonetheless, comparing the reaction rates of the different branching channels as a function of the gas temperature (using the modified Arrhenius equation, values in Table~\ref{tab:DR_reactions}), the DR channel forming CHD is the fastest, followed by the three-body channel forming HD, and then CH$_2$. While incomplete without accounting for destruction processes, a comparison of the reaction rates suggests that CHD is the preferred product of the DR of CH$_2$D$^+$ with electrons. In addition to the fast ion--molecule exchange reactions discussed above, neutral--neutral reactions involving deuterium exchange via either deuterium atoms or HD also form effective pathways for deuteration. The significance of such reactions was first proposed by \citet{Croswell1985} for the formation of OD in dark clouds (OH + D $\rightleftharpoons$ OD + H) and was more recently illustrated for hydrocarbon radicals by \citet{Dias2022}. Lacking laboratory measurements (to the best of our knowledge), the equivalent reaction forming CD (CH + D $\rightleftharpoons$ CD + H) is assumed to be extremely slow (or not to occur at all), which is due to the electronic configuration of carbon and the nature of the C$-$H bond, unlike in the case of oxygen and OH \citep{Kostyukevich2018}. Therefore, in the analysis that follows, we only evaluate the role of direct exchange reactions between CH and HD, which has a forward reaction rate of 1.1$\times10^{-10}$~s$^{-1}$ \citep[at $T = 40~$K;][]{Brownsword1997} in the formation of CD.

The lower values of deuteration derived from the CD/CH abundance ratio relative to those derived for other small molecules such as C$_2$H or H$_2$CO suggest that the former traces deuteration in more dilute ($\sim10^{3}\,$cm$^{-3}$) but warm gas layers in which CH$_2$D$^+$ is not yet the dominant channel for deuteration. In addition, as gas temperatures increase, DR reactions involving C$_2$HD$^+$ compete with that of CH$_2$D$^+$ to deuterate small hydrocarbons such as C$_2$H \citep{Millar1989}. This makes the CD/CH ratio a potentially sensitive tracer for the gas temperatures in such intermediate phases.

\begin{figure*}
    \centering
\includegraphics[width=0.98\textwidth]{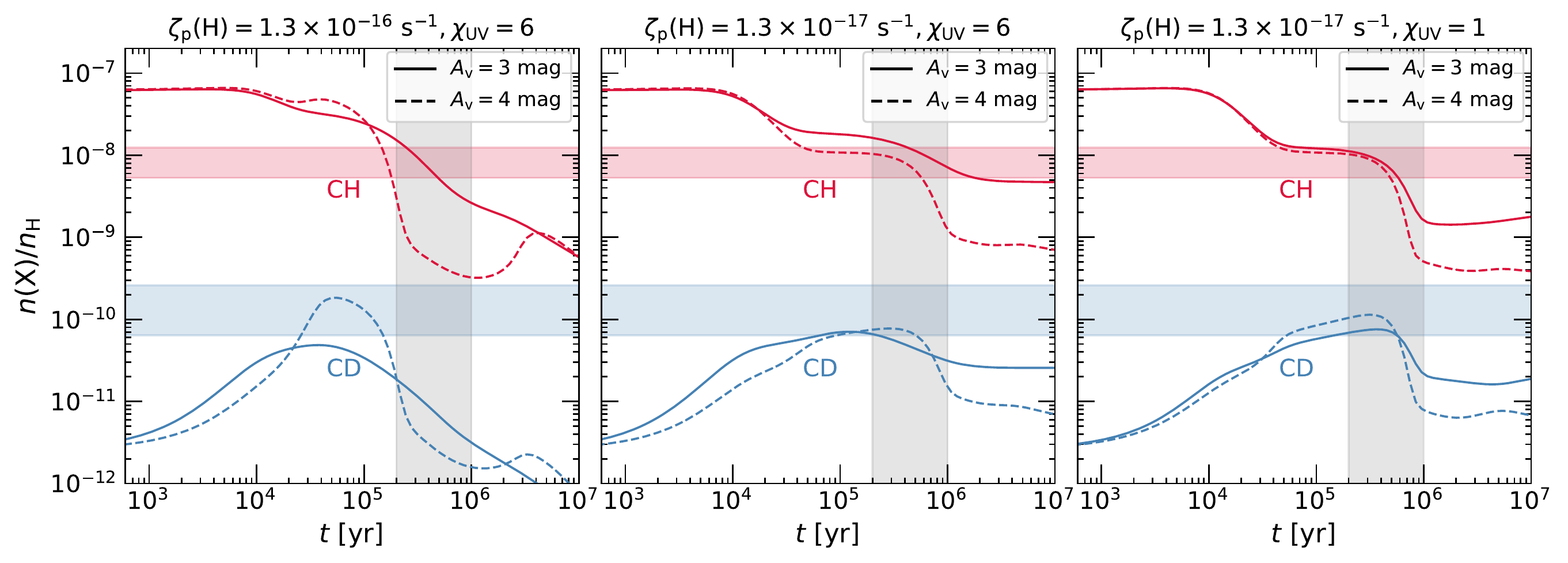}
    \caption{Fractional abundances reproduced using PyRate. From left to right: Fractional abundances of CH (in red) and CD (in blue) with respect to the total gas density ($n_{\rm H}$) for models with $\zeta_{\rm p}(\rm H)$ and $\chi_{\rm UV}$ = ($1.3\times10^{-16}~$s$^{-1}$, 6), ($1.3\times10^{-17}~$s$^{-1}$, 6), and ($1.3\times10^{-17}~$s$^{-1}$, 1), respectively. The solid curves and dashed curves represent the modelled abundances for $A_{\rm v} = 3$ and 4~mag, respectively, while the red and blue shaded regions represent the range of CH and CD abundances, respectively, as determined from the derived column densities. The grey shaded region represents the age of the source. }
    \label{fig:fractional_abundances}
\end{figure*}
To explore the physics at work and to evaluate the relative contributions of the different formation pathways of CD discussed above (see also Fig.~\ref{fig:network}), we carried out chemical simulations using the gas-grain chemical code pyRate described in \citet{Sipila2015, Sipila2019}. 

\subsection{The chemical model}\label{subsec:chemical_model}
The chemical code pyRate has been widely used to study deuterium chemistry and uses a chemical network adapted from the latest public release of the KIDA gas-phase network \citep{Wakelam2015} as well as the relevant photochemistry and has been benchmarked against other pseudo-time dependent chemical codes \citep{Sipila2015}. The models were run to explore the impact of varied physical conditions on the resulting abundances by assuming an initially atomic gas composition for all species except for hydrogen and deuterium, which are locked into H$_2$ and HD, respectively, where the initial abundances of different species are tabulated in Table~\ref{tab:initial_abundances}. The models were run assuming an average gas temperature of 30~K and primary cosmic-ray ionisation rates per H nucleon, $\zeta_{\rm p}({\rm H})$, of $1.3\times10^{-16}$ and $1.3\times10^{-17}$~s$^{-1}$, values that are typical for diffuse and translucent molecular clouds \citep{vanderTak2000, Indriolo2012, Neufeld2017}. The gas density, $n_{\rm H}$ and UV radiation field,
$\chi_{\rm UV}$, are fixed to $2\times10^{3}$~cm$^{-3}$ and 6 in Draine units based on constraints from observations of C$^{+}$ by \citet{Ceccarelli1998}. This assumption is valid given that CH is chemically associated
with gas layers traced by C$^+$, even showing line widths and velocities that are comparable. Furthermore, the models were run for visual extinctions, $A_{\rm V}$, of 3~mag and 4~mag computed using the average $A_{\rm V}$--$n_{\rm H}$ relation described in \citet{Bisbas2019} ($A_{\rm V}  = 0.05 {\rm exp}\left[1.6 \left(n_{\rm H}/[{\rm cm}^{-3}]\right)^{0.12}\right] \, [{\rm mag}]$), taking into account variations in the relation itself and uncertainties in the
assumed gas density.

Figure~\ref{fig:fractional_abundances} presents the modelled fractional abundances of CH and CD as a function of time, assuming a range of ${N_{\rm H}}$ values set by ${\sim 2\times10^{21} \left(A_{\upsilon}/[{\rm mag}] \right)~{\rm cm}^{-2}}$ 
\citep{Zhu2017} \citep[consistent with the values used in][]{Bottinelli2014B} and ${N_{\rm H} = 1.5\times10^{22}~{\rm cm}^{-2}}$ derived by \citet{vanDishoeck1995} for the low-density ambient cloud surrounding the protostar. The model that best reproduces the CH and CD abundances determined from our observations is that with $A_{\rm v}=4$~mag and $\zeta_{\rm p}(\rm H) = 1.3\times10^{-17}~$s$^{-1}$. The resulting value for $\zeta_{\rm p}(\rm H)$ is consistent with the values for the primary cosmic-ray ionisation rate derived in translucent gas layers traced by species such as H$_3^+$ \citep{Indriolo2012, Neufeld2017}. We note that recent Fermi-LAT observations present evidence for an excess cosmic-ray flux in the Ophiuchus region \citep{Baghmanyan2020}.  
This is because the cosmic-ray particles that are responsible for initiating astrochemical pathways via the ionisation of atomic and molecular hydrogen have low-energy ($<1$~GeV) particles and their flux is attenuated as they propagate into denser cloud environments. In contrast, Fermi-LAT measures diffuse $\gamma$-ray emission that originates from $\pi^{0}$-decay interactions of cosmic rays in the dense molecular clouds with energies of between a few GeV to the TeV range. Moreover, the excess cosmic-ray flux reported by \citet{Baghmanyan2020} reflects an enhancement in the cosmic-ray energy density spectrum with respect to the average spectral shape predicted by the Alpha Magnetic Spectrometer (AMS-02) for energies greater than 30~GeV.

To verify the validity of the modelled parameters, we also cross-check the fractional abundances reproduced by the same physical conditions for HCO$^+$ ---a species that is chemically associated with CH--- and its deuterated counterpart, DCO$^+$, in Fig.~\ref{fig:Fractional_abundance_HCOp}. Albeit not without uncertainties, because the HCO$^+$ molecular line emission probes slightly denser and cooler environments than the CH absorption, the abundances of HCO$^+$ and DCO$^+$ reproduced in our models are in general agreement with values derived by \citet{vanDishoeck1995} for models with $\zeta_{\rm p}({\rm H}) = 1.3\times10^{-17}~$s$^{-1}$ and $A_{\rm v} =4$~mag at times closer to the age of the protostellar envelope \citep[$t\sim$ a few $\times 10^{5}$ to $10^{6}~$years;][]{Brunken2014}. We note that the fractional abundance of HCO$^+$ is determined using the $N$(HCO$^+$) value reported in \citet{vanDishoeck1995} (see Table~8) for the ambient cloud component, while $N$(DCO$^+$) is estimated by scaling $N$(HCO$^+$) with the D/H ratio, assuming that the D/H ratio derived from HCO$^+$ is a constant across the protostellar envelope and surrounding gas, with a value of 0.0086.

\begin{figure}
     \centering  \includegraphics[width=0.45\textwidth]{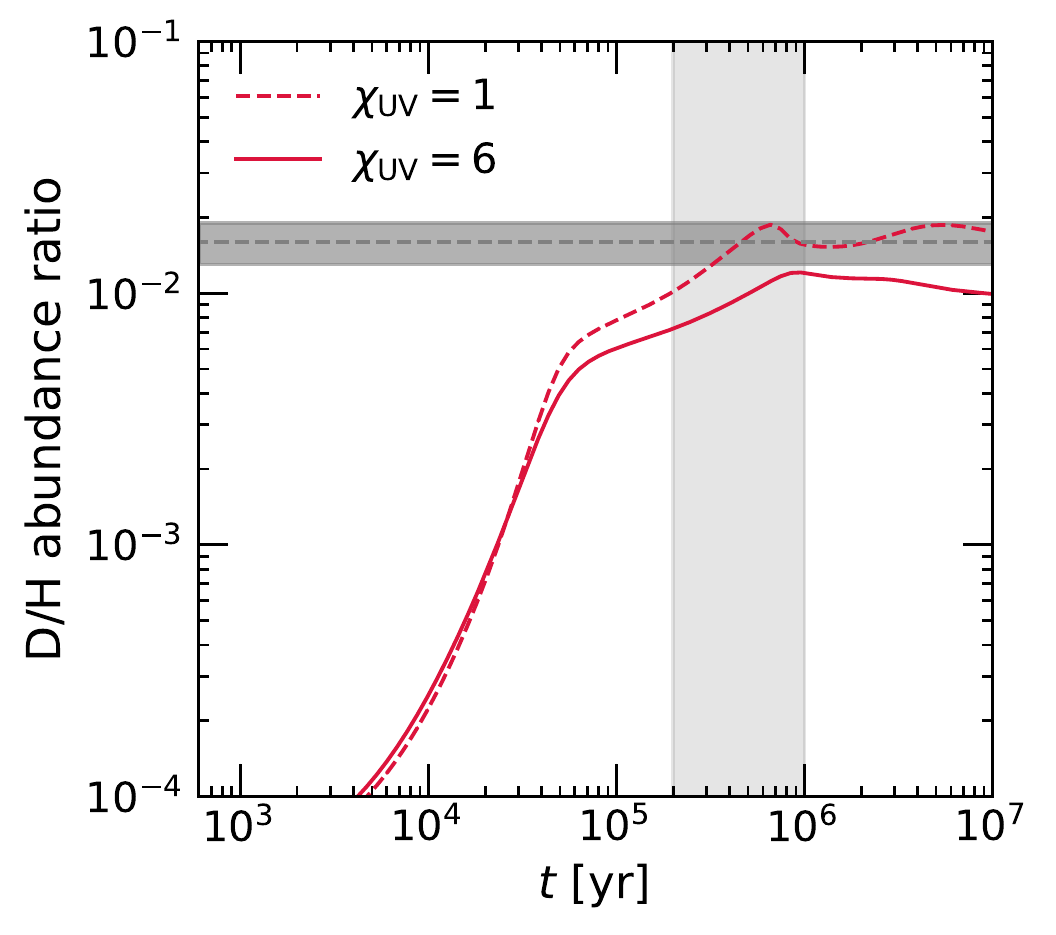}
     \caption{Modelled D/H abundance ratio obtained from CH/CD for models with $\zeta_{\rm p}({\rm H}) = 1.3\times10^{-17}$~s$^{-1}$, $A_{\rm v}=4~$mag, and $\chi_{\rm UV} = 1$ (dashed curve) and 6 (solid curve), where the horizontal and vertical grey shaded regions mark the D/H ratio derived from observations and the estimated age range of the system, respectively.}
     \label{fig:d_h_ratio}
 \end{figure}

By fixing $\zeta_{\rm p}({\rm H}) = 1.3\times10^{-17}~$s$^{-1}$ and $T=30~$K as the primary conditions for CH and CD, we then investigated the effects of the UV radiation field on the abundance by rerunning the models for the standard ISM value of $\chi_{\rm UV} = 1$. We find that at times, $t$, equivalent to the age of the system, models that were run at both $A_{\rm v} = 3$ and 4~mag match the abundances of CH rather well, while those of CD are better represented by models with $A_{\rm v} = 4~$mag. Therefore, accounting for the entire range of D/H ratios determined from the observations of CD and CH (see Fig.~\ref{fig:d_h_ratio}), the subsequent analysis is carried out for models with $\zeta_{\rm p}({\rm H}) = 1.3\times10^{-17}$~s$^{-1}$, $A_{\rm v}=4~$mag, and $\chi_{\rm UV} = 1$ and 6.

\subsection{Temperature dependence of the CD/CH ratio}
\begin{figure*}
    \sidecaption
\includegraphics[width=12cm]{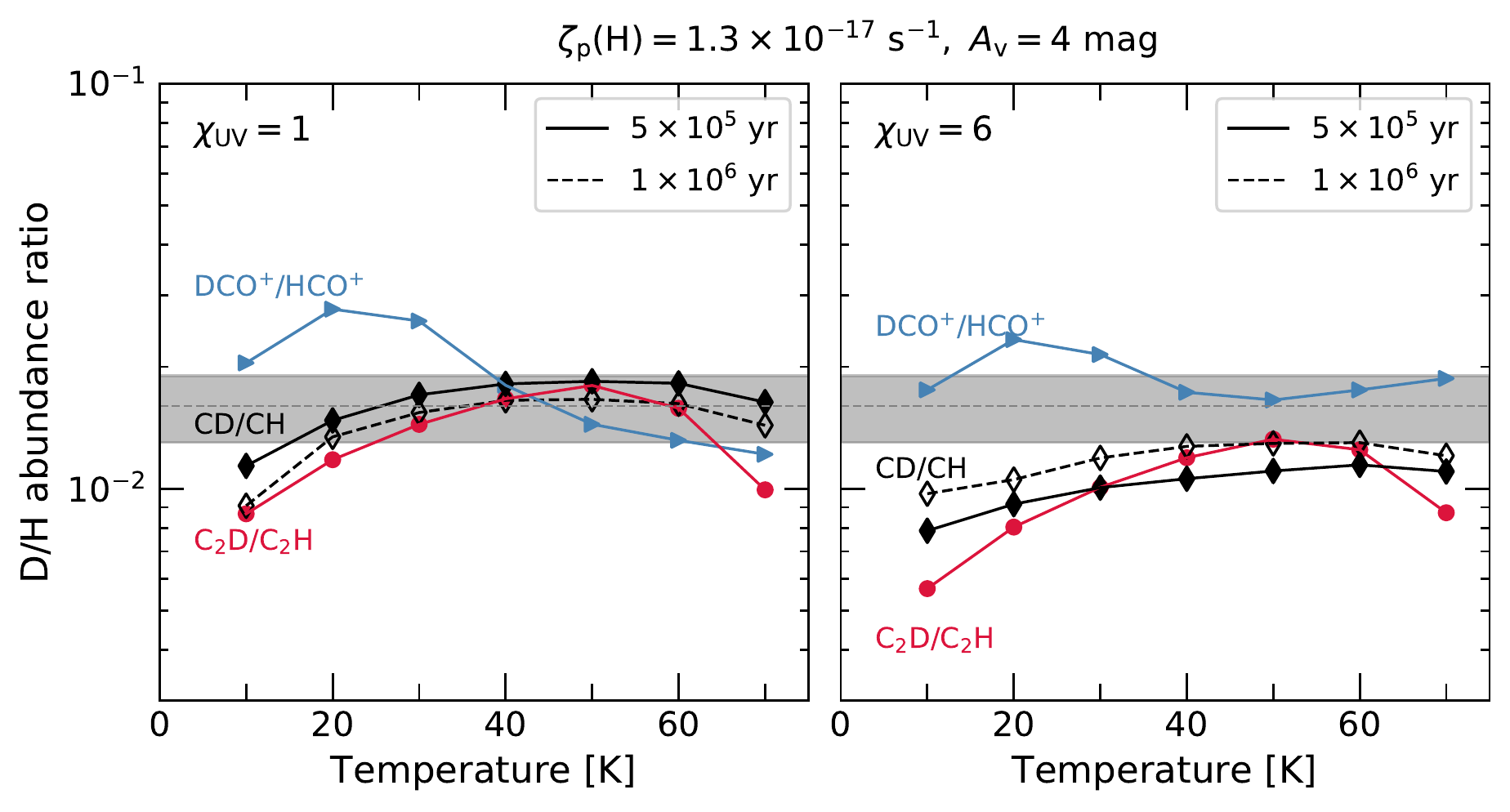}
    \caption{Modelled D/H abundance ratio for CH as a function of temperature computed for times $t=5\times10^{5}~$years (solid curve with diamond markers) and $1\times10^{6}~$years (dashed curve) for models with $\chi_{\rm UV} =1$ (left) and $\chi_{\rm UV}=6$ (right), where the horizontal grey shaded regions mark the dispersion in the CD/CH abundance ratio derived toward IRAS\,16293$-$2422. For comparison, we also display the temperature dependence on the D/H abundance ratio for HCO$^+$ (blue triangles) and C$_2$H (red circles) at the time of $t=5\times10^{5}~$years.}
    \label{fig:temperature_dependence_models}
\end{figure*}

Figure~\ref{fig:temperature_dependence_models} explores the temperature dependence of the CD/CH abundance ratio at simulation times, $t=5\times10^{5}$ and $1\times10^{6}$~years using the physical conditions of the best-fit model. While models exposed to higher values of $\chi_{\rm UV}$ take longer to attain the same abundance ratio owing to the increased photodissociation of CH, the general trends at low to moderate temperatures are similar. For both models, as the temperature increases from 10~K to 20~K and 20~K to 30~K, the CD/CH abundance ratio steadily increases, with a steeper slope at later times, reflecting the effectiveness of deuteration by CH$_2$D$^+$, as H$_2$D$^+$ is increasingly destroyed by reactions with H$_2$.

At $t=5\times 10^{5}$~years and $1\times10^{6}$~years, for temperatures above 30~K in models with $\chi_{\rm UV} = 1$, the degree of deuteration in CH increases monotonically until $T = 50~$K, beyond which it decreases, showing a more rapid drop after $T = 60~$K. Models exposed to a higher UV radiation field ($\chi_{\rm UV} = 6$) on the other hand show a much flatter temperature dependence, decreasing only slightly after the threshold temperature of 60~K, with a more prominent temperature dependence seen at simulation times of $t = 5\times10^{5}~$years. The models best reproduce the derived CD/CH ratio at gas temperatures of between 20 and 30~K at a simulation time of $5\times10^{5}$~years and between 30 and 60~K at a simulation time of $10^{6}$~years for the model with $\chi_{\rm UV}=1$. This suggests that the CD abundance peaks in gas layers where deuterium fractionation is driven by reactions with CH$_2$D$^+$. This is evident from the temperature dependence of DCO$^+$/HCO$^+$ (also shown in Fig.~\ref{fig:temperature_dependence_models}), which peaks at 20~K, above which the ratio decreases because the deuteration in this species is mainly driven by reactions between H$_2$D$^+$ and CO for the physical conditions modelled (see Fig.~\ref{fig:network}). In contrast, the deuteration in C$_2$H follows a similar trend to that of CH, but with the ratio peaking at temperatures close to 50~K as discussed in \citet{Roueff2007}. This suggests that the CD/CH abundance ratio is likely a tracer of deuterated chemistry at intermediate, `warm' gas layers. However, while the CD/CH abundance ratio shows a penchant for higher gas temperatures ($\geq30~$K) as discussed above, it is difficult to estimate the temperature of the low-density layers of the envelope surrounding IRAS\,16293$-$2422 for certain, without more robust constraints on the other modelled parameters.

\subsection{Formation routes of CD}
Finally, for the physical conditions discussed above, we examine the formation routes of CD for gas temperatures fixed to 20~K and 40~K in order to explore the ensuing chemistry in cold and warm gas, respectively (Fig.~\ref{fig:formation_routes}). The dominant formation pathway for CD is the DR of CH$_2$D$^+$ with electrons for both sets of models (those with $\chi_{\rm UV} = 1$ and $\chi_{\rm UV} = 6$) across the different temperatures. This is followed by neutral-exchange reactions between CHD and atomic hydrogen for models with $\chi_{\rm UV}=1$, whereas neutral exchange reactions between CH$_2$ and deuterium atoms become competitive if not dominant  in models with $\chi_{\rm UV}=6$ . While contributions to the observed abundances of CD via the DR of C$_2$HD$^+$ with electrons are more significant at $t<10^{5}~$years, the DR of CHD$_2^+$ with electrons becomes relatively more important at later times. Unsurprisingly, in models exposed to a higher UV radiation field, photodissociation reactions play a more prevalent role.    
Therefore, for the conditions modelled here, the main formation pathway for CD is via the DR of CH$_2$D$^+$ and not via direct neutral-exchange reactions. This supports our hypothesis that CD is indeed an important tracer for deuterium chemistry initiated by reactions involving CH$_2$D$^+$.

\subsection{Dependence of the CD/CH ratio on the initial ortho-to-para ratio of \texorpdfstring{H$_2$}{H2}}

The simulations described above were run assuming an initial ortho-to-para ratio (OPR) for H$_2$ of 10$^{-3}$, which is motivated by the results of recent 3D magnetohydrodynamic simulations of molecular clouds by \citet{Lupi2021}, who investigated the sensitivity of this ratio on gas densities. The models by \citet{Lupi2021} predict low values for the OPR as the gas becomes fully molecular, with values between 10$^{-3}$ and 10$^{-2}$ at moderate gas densities in the range of 10$^3$--10$^4$~cm$^{-3}$ (corresponding to the gas densities modelled in this work). The value of the initially assumed OPR of H$_2$ not only influences the populations of the
nuclear spin states of H$_3^+$ and its deuterated isotopologues, but also the degree of deuteration in subsequently formed species. As the initial
OPR of H$_2$ increases, the deuteration of H$_3^+$ is inhibited because the presence of ortho-H$_2$ allows backward processes (in the proton-deuterium exchange reaction) to proceed efficiently \citep[see][]{Flower2006}. On the contrary, assuming low values for the initial OPR of H$_2$ can boost the abundance of H$_2$D$^+$ in the simulations. Therefore, we examine the dependence of the fractional abundances of CH and CD on the value of the H$_2$ OPR used in the models by carrying out additional simulations for initial OPRs of H$_2$ of 0.1 and 3.0. Figure~\ref{fig:fractional_abundances_OPR} is similar to Fig.~\ref{fig:fractional_abundances}  and displays the fractional abundances of CH and CD as a function of time but for simulations run with an OPR for H$_2$ = 0.1 (top panel) and 3.0 (bottom panel). We find variations in the abundances of both CH and CD to be marginal and to result in deviations in the subsequently determined D/H ratio of at most 5\,\% and 7\,\% for those models with an OPR of H$_2$ = 0.1 and 3.0, respectively (see Fig.~\ref{fig:temperature_dependence_OPR}). This result is unsurprising, given that deuteration in CH is driven by CH$_2$D$^+$ and not H$_2$D$^+$. Varying the OPR of H$_2$ in the models decreases the D/H ratio of C$_2$H by ${\sim\!9\,\%}$, while that of HCO$^+$ decreases by ${\sim\!55\,\%}$ and 68\,\% at lower gas temperatures for models with H$_2$ OPR = 0.1 and 3.0, respectively, approaching the same value as the temperature increases beyond 60~K. This difference is a direct reflection of the fact that deuteration in HCO$^+$ stems from H$_2$D$^+$ at low temperatures, the abundance of which varies with the OPR of H$_2$. Therefore, we conclude that the initial choice of OPR of H$_2$ in the simulations does not directly influence deuteration in CH.

\section{Conclusions} \label{sec:conclusions}
In this paper, we report the first detection of CD in
the ISM, which we detect towards IRAS\,16293$-$2422 in absorption via observations of the HFS splitting lines of both $\Lambda$-doublets of the $N=1, J=3/2\rightarrow1/2$ transition made with the APEX 12~m telescope. The detection of CD provides a new opportunity to observationally probe deuteration proceeding in warm gas via reaction channels involving CH$_2$D$^+$. By combining the column density of CD derived from our observation with that from the corresponding transitions of CH towards the same source, we estimate a D/H ratio of $0.016 \pm 0.03$. Remarkably, compared with other species that inherit deuterium from CH$_2$D$^+$, the degree of fractionation derived from CH is lower by an order of magnitude. 
This could likely be due to the fact that the CD/CH abundance ratio traces the onset of deuterium chemistry initiated via reactions with CH$_2$D$^+$, unlike in the case of C$_2$D and HDCO whose abundances are enhanced as reactions driven by CH$_2$D$^+$ become the predominant chemical pathway for deuteration. Furthermore, astrochemical model calculations carried out provide evidence that deuteration in CH proceeds via `warm deuterium chemistry', with the models best reproducing the observed ratio at gas temperatures of $>30$~K. However, better constraints, particularly on the physical parameters of the cloud, are necessary to ascertain the role of CD as a probe of warm deuterium chemistry and the possible use of the CD/CH ratio in constraining envelope gas temperatures. Coordinated laboratory efforts in the future ---measuring frequencies of sister species of CD, CHD and CD$_2$, and refining the values of relevant reaction rates that would allow determinations of the branching ratios for DR reactions involving deuterated isotopes of CH$_3^+$--- will make it possible to improve our understanding of the deuterium chemistry at work in moderately warm environments of the ISM. Certainly, the successful detection of CD in IRAS\,16293$-$2422 provides new evidence for deuterated chemistry initiated by CH$_2$D$^+$.

\begin{acknowledgements}
This publication is based on data acquired with the Atacama Pathfinder Experiment (APEX) under the project id:  M9530C$\_$107. APEX is a collaboration between the Max-Planck-Institut fur Radioastronomie, the European Southern Observatory, and the Onsala Space Observatory. We would like to express our gratitude to the APEX staff and science team for their continued assistance in carrying out the observations presented in this work. Part of this work reports observations made with the NASA/DLR Stratospheric Observatory for Infrared Astronomy (SOFIA). SOFIA is jointly operated by the Universities Space Research Association, Inc. (USRA), under NASA contract NNA17BF53C, and the Deutsches SOFIA Institut (DSI) under DLR contract 50 OK 0901 to the University of Stuttgart. A.M.J. acknowledges support by USRA through a grant for SOFIA Program 08-0038. O.S. would like to acknowledge financial support from the Max Planck Society. We would like to thank the anonymous referee for valuable comments as well as Antonio Hernandez-Gomez, David Neufeld, Jacques Le Bourlot and Jamila Pegues for their helpful discussions on the source IRAS\,16293$-$2422 and the use of varied chemical modeling tools. The authors would like to express our gratitude to the developers of the many Python libraries, made available as open-source software, in particular this research has made use of the NumPy \citep{numpy}, SciPy \citep{scipy} and matplotlib \citep{matplotlib} packages.  
\end{acknowledgements}

\bibliographystyle{aa} 
\bibliography{ref}

\begin{appendix}
\section{Additional information}\label{appendix:additional_info}

\begin{table}[h]
    \centering
    \caption{Spectroscopic parameters of the $N=2 \rightarrow 1, J=3/2\rightarrow1/2$ transitions of CD studied using SOFIA/4GREAT. }
    \begin{tabular}{llrcl}
    \hline \hline 
          \multicolumn{2}{c}{Transition}& \multicolumn{1}{c}{Frequency} & \multicolumn{1}{l}{$A_{\rm ul}\times10^{-3}$} 
          \\
          Parity & $F^{\prime} - F^{\prime \prime}$ & \multicolumn{1}{c}{[MHz]} &  \multicolumn{1}{c}{[s$^{-1}$]} & \\
         \hline
     $- \rightarrow +$ & $1/2 \rightarrow 3/2$ & 1325248.368(276) &  0.61  
     \\
                      & $1/2 \rightarrow 1/2$ & 1325251.364(278) &  4.87 
                      \\
                       & $3/2 \rightarrow 3/2$ & 1325255.172(210) &  2.43  
                       \\
                       & $3/2 \rightarrow 1/2$ & 1325258.167(213) &  3.04  
                       \\
                       & $5/2 \rightarrow 3/2$ & 1325266.217(194) &  5.48  
                       \\
     $+ \rightarrow -$ & $1/2 \rightarrow 3/2$ & 1325775.667(261) &  0.61  
     \\
                       & $3/2 \rightarrow 3/2$ & 1325784.078(197) &  2.45  
                       \\
                       & $1/2 \rightarrow 1/2$ & 1325792.765(258) &  4.90 
                       \\
                       & $5/2 \rightarrow 3/2$ & 1325797.812(194) &  5.52  
                       \\
                       & $3/2 \rightarrow 1/2$ & 1325801.181(195) &  3.06  
                       \\
  
\hline
    \end{tabular}
\tablefoot{The frequencies and other spectroscopic parameters are taken from the Jet Propulsion Laboratory \citep[JPL;][]{Pickett1998} database. Values in parenthesis represent uncertainties in the rest frequencies of the HFS lines, in units of the last significant digits.}
    \label{tab:sofia_freq}
\end{table}

\begin{table}[h]
    \centering
    \caption{Spectroscopic parameters of the $N = 1, J=3/2 \rightarrow 1/2$ transitions of the CH discussed in this study. }
    \begin{tabular}{lllcl}
    \hline \hline 
          \multicolumn{2}{c}{Transition}& \multicolumn{1}{c}{Frequency} & \multicolumn{1}{l}{$A_{\rm ul}\times10^{-4}$} 
          \\
          Parity & $F^{\prime} - F^{\prime \prime}$ & \multicolumn{1}{c}{[MHz]} &  \multicolumn{1}{c}{[s$^{-1}$]} & \\
         \hline
     $- \rightarrow +$ & $1 \rightarrow 1$ & 532721.588(2) &  2.1  
     \\
                      & $ 2 \rightarrow 1$\tablefootmark{*} & 532723.889(1) &  6.2  
                      \\
                       & $1 \rightarrow 0$ & 532793.274(1) &  4.1  
                       \\
     $+ \rightarrow -$ & $2 \rightarrow 1$\tablefootmark{*} & 536761.046(1) &  6.4  
     \\
                       & $1 \rightarrow 1$ & 536781.856(1) &  2.1  
                       \\
                       & $1 \rightarrow 0$ & 536795.569(1) &  4.3  
                       \\
  
\hline
    \end{tabular}
    \tablefoot{{The frequencies and other spectroscopic parameters are taken from \citet{Truppe2014} and the Cologne Database for Molecular Spectroscopy \citep[CDMS;][]{Endres2016}. Values in parenthesis represent uncertainties in the rest frequencies of the HFS lines, in units of the last significant digits.} \tablefoottext{*}{Indicates the HFS splitting transition, which was used to set the velocity scale in the analysis.}}
    \label{tab:ch_freq}
\end{table}

\begin{table}[]
\centering
    \caption{Reaction rates of the different dissociative recombination channels of CH$_2$D$^+$ with an electron.}
    \begin{tabular}{lcccc}
    \hline \hline
    \multicolumn{1}{c}{Product} & $\alpha$ & $\beta$ & $\gamma$ & $k$\\
     (CH$_2$D$^+$ + $e^- \rightarrow$)       & [cm$^3$~s$^{-1}$] & &  & [cm$^3$~s$^{-1}$]\\
    \hline
     C + H + HD & $2.0\times10^{-7}$ & $-0.3$ & 0.0  & $4.0\times10^{-7}$ \\
     C + D + $p$-H$_2$    & $5.0\times10^{-8}$ & $-0.3$ & 0.0 & $1.0\times10^{-7}$\\
     C + D + $o$-H$_2$    & $5.0\times10^{-8}$ & $-0.3$ & 0.0 & $1.0\times10^{-7}$ \\
     H + H + CD & $5.3\times10^{-8}$ & $-0.3$ & 0.0 & $1.1\times10^{-7}$  \\
     H + D + CH    & $1.1\times10^{-7}$ & $-0.3$ & 0.0 & $2.2\times10^{-7}$ \\
     CH + HD    & $9.3\times10^{-8}$ & $-0.3$ & 0.0 & $1.8\times10^{-7}$ \\
     CD + $p$-H$_2$    & $2.3\times10^{-8}$ & $-0.3$ & 0.0 & $4.6\times10^{-8}$ \\
     CD + $o$-H$_2$    & $2.3\times10^{-8}$ & $-0.3$ & 0.0 & $4.6\times10^{-8}$\\
     H + CHD & $2.7\times10^{-7}$ & $-0.3$ & 0.0  & $5.4\times10^{-7}$ \\
     D + CH$_2$    & $1.3\times10^{-7}$ & $-0.3$ & 0.0 & $2.6\times10^{-7}$\\
     CH$_2$D + photon    & $1.0\times10^{-10}$ & $-0.7$ & 0.0 & $5.0\times10^{-10}$ \\
     \hline
    \end{tabular}
    \tablefoot{Left to right: $\alpha$, $\beta$ and $\gamma$ represent the pre-exponential factor and power indices, respectively, where their values are taken from \citet{Sipila2015} and \cite{Sipila2019}. The reaction rates, $k$, are computed for a gas temperature of 30~K following the modified Arrhenius equation, $k(T) = \alpha\left(T/300\right)^{\beta} e^{-\gamma/T}$.}
    \label{tab:DR_reactions}
\end{table}

\begin{table}[]
    \centering
    \caption{Initial abundances used in the chemical models.}
    \begin{tabular}{lr}
    \hline \hline
    Species & Abundance ($n({\rm X})$/$n_{\rm H}$) \\
   & \\
    \hline
$p$-H$_2$   & $4.99\times10^{-1}$ \\
$o$-H$_2$  & $5.00\times10^{-4}$ \\
HD    & $1.60\times10^{-5}$\\
He    & $9.00\times10^{-2}$ \\
C$^+$    & $1.20\times10^{-4}$\\
N     & $7.60\times10^{-5}$\\
O     & $2.56\times10^{-4}$\\
F     & $2.00\times10^{-9}$ \\
S$^+$    & $8.00\times10^{-8}$\\
Si$^+$   & $8.00\times10^{-9}$\\
Na$^+$   & $2.00\times10^{-9}$\\
Mg$^+$   & $7.00\times10^{-9}$\\
Fe$^+$   & $3.00\times10^{-9}$\\
P$^+$    & $2.00\times10^{-10}$\\
Cl$^+$   & $1.00\times10^{-9}$\\
\hline
    \end{tabular}
      \label{tab:initial_abundances}
\end{table}

\begin{figure*}
\sidecaption
\includegraphics[width=12cm]{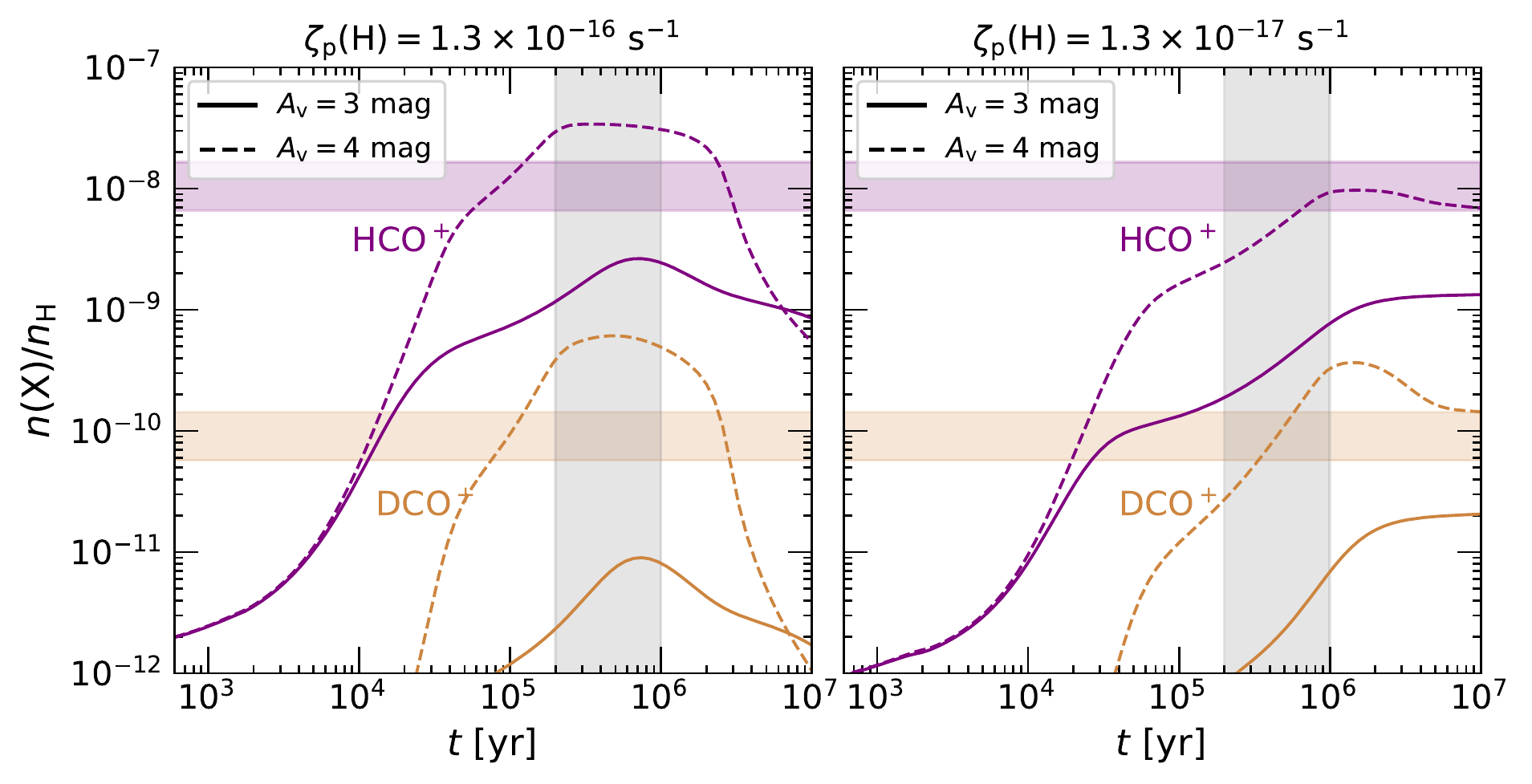}
  \caption{Modelled fractional abundances of HCO$^+$ (in purple) and DCO$^+$ (in yellow) with respect to the total gas density ($n_{\rm H}$) for models with $\zeta_{\rm p}(\rm H)$ and $\chi_{\rm UV}$ = ($1.3\times10^{-16}~$s$^{-1}$, 6) (left) and  ($1.3\times10^{-17}~$s$^{-1}$, 6) (right), respectively. The solid and dashed curves represent the fractional abundances for $A_{\rm v} = 3$ and 4~mag, respectively, while the purple and yellow shaded regions represent the range of HCO$^+$ and DCO$^+$ abundances, respectively, as determined from values reported in \citet{vanDishoeck1995}. The vertical grey shaded region represents the approximate age of the source. }
    \label{fig:Fractional_abundance_HCOp}
\end{figure*}

\begin{figure*}
    \centering
\includegraphics[width=1\textwidth]{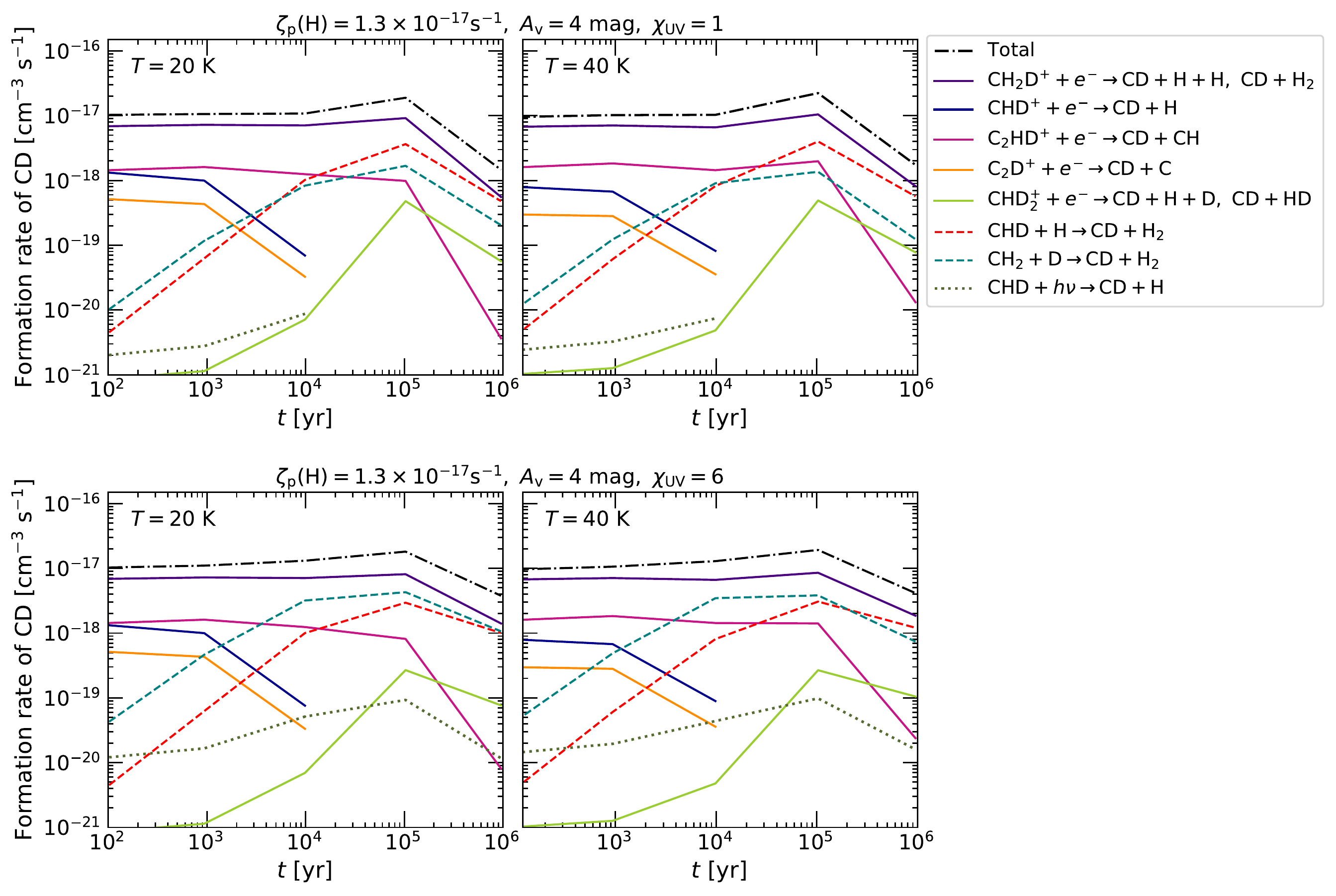}
    \caption{Comparison of the most relevant reaction pathways for the formation of CD as a function of time for models with $\chi_{\rm UV}=1$ (top panel) and $\chi_{\rm UV}=6$ (bottom panel) at $T = 20$~K (left) and $T = 40~$~K (right). Curves representing the DR rates for CHD$^+$ and C$_2$D$^+$ stop at simulation times of $t=10^{4}~$years because the rates for these reactions significantly decrease beyond this point. }   \label{fig:formation_routes}
\end{figure*}

\begin{figure*}
    \centering
    \includegraphics[width=0.9\textwidth]{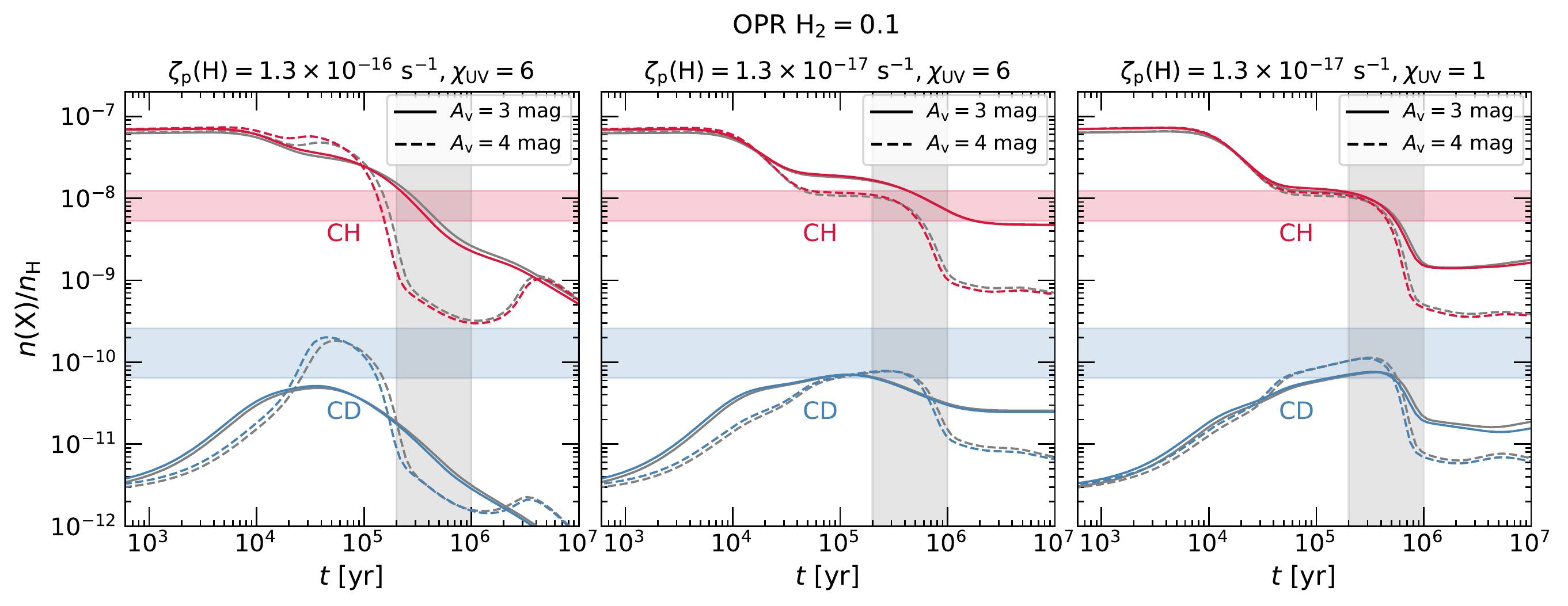}\quad
    \includegraphics[width=0.9\textwidth]{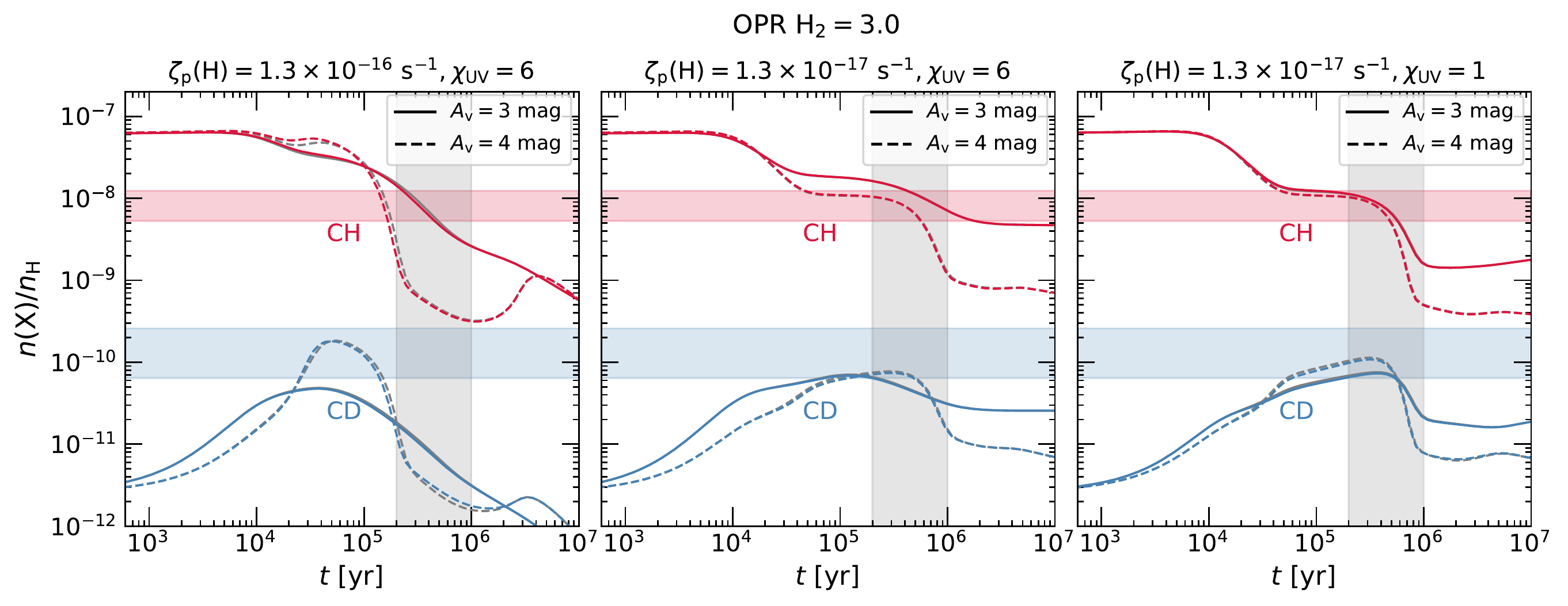}
    \caption{Same as Fig.~\ref{fig:fractional_abundances} but for models with an initial OPR of H$_2$ = 0.1 (top panel) and 3.0 (bottom panel). For comparison, the grey solid and dashed curves present the results obtained from models with an OPR of H$_2$ =10$^{-3}$ and $A_{\rm v} =3$ and 4~mag, respectively. }
    \label{fig:fractional_abundances_OPR}
\end{figure*}

\begin{figure}
    \includegraphics[width=0.6\textwidth]{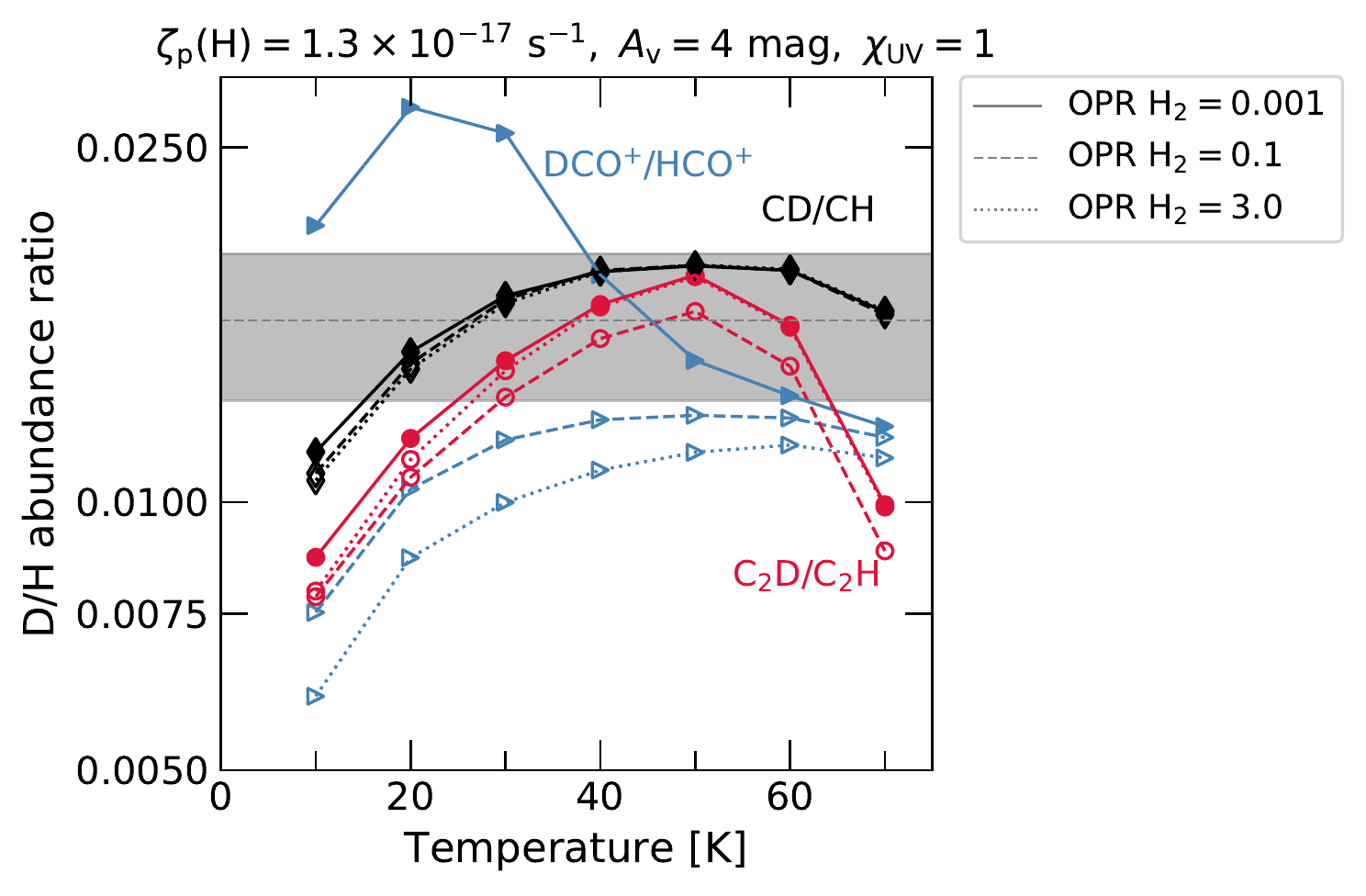}
    \caption{Modelled D/H abundance ratio for CH (black diamonds), HCO$^+$ (blue triangles) and C$_2$H (red circles) as a function of temperature computed for a time of $t = 5\times10^{5}$~years, where the horizontal grey shaded regions mark the dispersion in the CD/CH abundance ratio derived toward IRAS~16293$-$2422. The solid, dashed, and dotted curves represent simulations run for OPRs of H$_2$ = 10$^{-3}$, 0.1, and 3.0, respectively.}
\label{fig:temperature_dependence_OPR}
\end{figure}
\end{appendix}
\end{document}